\documentclass[pra,twocolumn,showpacs,superscriptaddress,nofootinbib]{revtex4-2}
\usepackage{graphicx}
\usepackage{braket}
\usepackage{amsmath,amsfonts,amssymb,amstext,amscd,amsthm,bbold}
\usepackage{comment,float}
\usepackage[dvipsnames]{xcolor}
\definecolor{myblue}{HTML}{03045E}
\usepackage[colorlinks=true,linkcolor=blue,urlcolor=myblue,citecolor=myblue]{hyperref}
\usepackage[normalem]{ulem}
\usepackage{mathtools}

\usepackage{lipsum}
\usepackage{dcolumn}
\usepackage{bm}
\usepackage{verbatim}

\usepackage{orcidlink}
\hypersetup{
colorlinks=true,
citecolor= myblue,
linkcolor=myblue,
filecolor=myblue,      
urlcolor=myblue,
pdftitle={Overleaf Example},
pdfpagemode=FullScreen,
}

\usepackage[T1]{fontenc}
\usepackage[utf8]{inputenc}

\usepackage{lipsum} % Include the lipsum package

\makeatletter
\newcommand{\fmarki}{*}
\newcommand{\fmarkii}{\ensuremath{\dagger}}
\newcommand{\fmarkiii}{\ensuremath{\ddagger}}
\newcommand{\fmarkiv}{\ensuremath{\mathsection}}
\newcommand{\fmarkv}{\ensuremath{\mathparagraph}}
\newcommand{\fmarkvi}{\ensuremath{\|}}
\newcommand{\fmarkvii}{**}
\newcommand{\fmarkviii}{\ensuremath{\dagger\dagger}}
\newcommand{\fmarkix}{\ensuremath{\ddagger\ddagger}}

\def\@fnsymbol#1{{\ifcase#1\or \fmarki\or \fmarkii\or \fmarkiii\or \fmarkiv\or \fmarkv\or \fmarkvi\or \fmarkvii\or \fmarkviii\or \fmarkix \else\@ctrerr\fi}}
\makeatother

\renewcommand{\fmarki}{$\dagger$}
\renewcommand{\fmarkii}{b$_2$}
\renewcommand{\fmarkiii}{c$_3$}
\renewcommand{\fmarkiv}{a$_4$}
\renewcommand{\fmarkv}{x$_5$}
\renewcommand{\fmarkix}{z$_9$}
\makeatletter
\renewcommand\section{\@startsection{section}{1}{0pt}{-2ex}{1ex}{\raggedright\bfseries\textsf}}
\renewcommand\subsection{\@startsection{subsection}{2}{0pt}{-2ex}{1ex}{\raggedright\bfseries\textsf}}
\renewcommand\subsubsection{\@startsection{subsubsection}{2}{0pt}{-2ex}{1ex}{\raggedright\bfseries\textsf}}
\makeatother

\begin{document}

\title{\textsf{Information scrambling in quantum walks: Discrete-time formulation of Krylov complexity}}% Force line breaks with 
\author{Himanshu Sahu$^{\orcidlink{0000-0002-9522-6592}}$}
\email{hsahu@perimeterinstitute.ca}
\affiliation{Perimeter Institute for Theoretical Physics, Waterloo, Ontario, N2L 2Y5, Canada.}
\affiliation{Department of Physics and Astronomy and Institute for Quantum Computing,University of Waterloo, Ontario N2L 3G1, Canada.}
\affiliation{Department of Physics and Department of Instrumentation \& Applied Physics, Indian Institute of Science, Bangalore - 560012, Karnataka, India.}

\begin{abstract}
We study information scrambling --- a spread of initially localized quantum information into the system's many degree of freedom --- in discrete-time quantum walks. We consider out-of-time-ordered correlators (OTOC) and  K-complexity as a probe of information scrambling.  The OTOC for local spin operators in all directions has a light-cone structure which is ``shell-like''. As the wavefront passes, the OTOC approaches to zero in the long-time limit, showing no signature of scrambling. The introduction of spatial or temporal disorder changes the shape of the light-cone localization of wavefunction. We formulate the K-complexity in system with discrete-time evolution, and show that it grows linearly in discrete-time quantum walk. The presence of disorder modifies this growth to sub-linear. Our study present interesting case to explore many-body phenomenon in a discrete-time quantum walk using scrambling.
\end{abstract}
\maketitle

\section{Introduction}\label{sec:Introduction}
Quantum scrambling\,\cite{PRXQuantum.5.010201,lewis-swanDynamicsQuantumInformation2019} is the process where interactions within a quantum system spread local information across its many degree of freedom. It's a fundamental process behind how isolated quantum systems reach thermal equilibrium\,\cite{PhysRevA.43.2046,rigolThermalizationItsMechanism2008,PhysRevE.50.888}, and it's closely linked to quantum chaos\,\cite{maldacenaBoundChaos2016}, the black-hole information problem\,\cite{yasuhirosekinoFastScramblers2008a,lashkariFastScramblingConjecture2013a,Shenker:2013pqa}, and how disorder affects collective spins in many-body systems\,\cite{PhysRevLett.123.165902,PhysRevB.95.060201}. The concept of scrambling also lays the groundwork for developing algorithms in quantum benchmarking and machine learning, which can make the exploration of Hilbert spaces more efficient\,\cite{mcclean_barren_2018,haferkamp_efficient_2023,PhysRevLett.124.200504,PhysRevResearch.3.L032057,Garcia:2021bin}.

There remain ambiguity is describing the process of quantum scrambling. One method to probe it involves using an out-of-time-ordered correlator\,\cite{swingleUnscramblingPhysicsOutoftimeorder2018a,PRXQuantum.5.010201}. For systems that exhibit a semi-classical limit or have a large number of local degrees of freedom, this correlator shows exponential growth, which can be used to identify a quantum counterpart to the Lyapunov exponent (LE), thus linking it to classical chaos\,\cite{PhysRevLett.118.086801}. Another approach involves studying the evolution dynamics of operators in Krylov space. Here, operator growth is measured by the ``K-complexity'', indicating the extent of delocalization of initial local operators evolving under Heisenberg evolution under the system Hamiltonian\,\cite{PhysRevX.9.041017,barbonEvolutionOperatorComplexity2019a,rabinovici_operator_2021}. It is speculated that this K-complexity grows exponentially in most generic nonintegrable systems\,\cite{PhysRevX.9.041017}. This exponential growth in K-complexity can be used to extract LE, establishing a connection with out-of-time-ordered correlators\,\cite{maldacenaBoundChaos2016}. Recent studies have explored K-complexity in various systems such as Ising models\,\cite{rabinovici_krylov_2022,rabinoviciKrylovLocalizationSuppression2022,PhysRevD.109.066010}, Sachdev-Ye-Kitaev (SYK) models \cite{jianComplexityGrowthOperators2021,bhattacharjeeKrylovComplexityLarge2023,heQuantumChaosScrambling2022}, quantum field theories\,\cite{caputaOperatorGrowth2d2021,khetrapalChaosOperatorGrowth2023,kunduStateDependenceKrylov2023,camargoKrylovComplexityFree2023,avdoshkinKrylovComplexityQuantum2022, Erdmenger:2023wjg}, the many-body localization system\,\cite{10.21468/SciPostPhys.13.2.037, Bento:2023bjn}, and open quantum systems\,\cite{bhattacharyaOperatorGrowthKrylov2022b,bhattacharjeeOperatorGrowthOpen2023,bhattacharyaKrylovComplexityOpen2023d,PhysRevResearch.5.033085, Bhattacharjee:2023uwx}.

Experimentally tunable toy models serve as valuable tools for investigating different phenomena from theoretical physics. One such tool, which we will pursue  in this work, is the quantum walk --- a quantum version of the classical random walk\,\cite{venegas-andraca_quantum_2012,ambainis_quantum_2003}. Specifically, we examine discrete-time quantum walks, which were previously employed to simulate controlled dynamics in quantum systems\,\cite{oka_breakdown_2005,PhysRevLett.121.260501,PhysRevLett.124.050502} and to construct quantum algorithms\,\cite{ambainis_quantum_2003}. These quantum walks are implemented experimentally using both lattice-based quantum systems and circuit-based quantum processors\,\cite{manouchehriPhysicalImplementationQuantum2014a,doi:10.1126/science.1174436,doi:10.1126/science.1231440}. The adaptability of quantum walks, allowing for the experimental modeling of various phenomena like topological effects\,\cite{PhysRevLett.124.050502,wuTopologicalQuantumWalks2019a}, therefore, also positions them as a promising platform for studying scrambling.

\begin{figure}
	\centering
	\includegraphics[width=\linewidth]{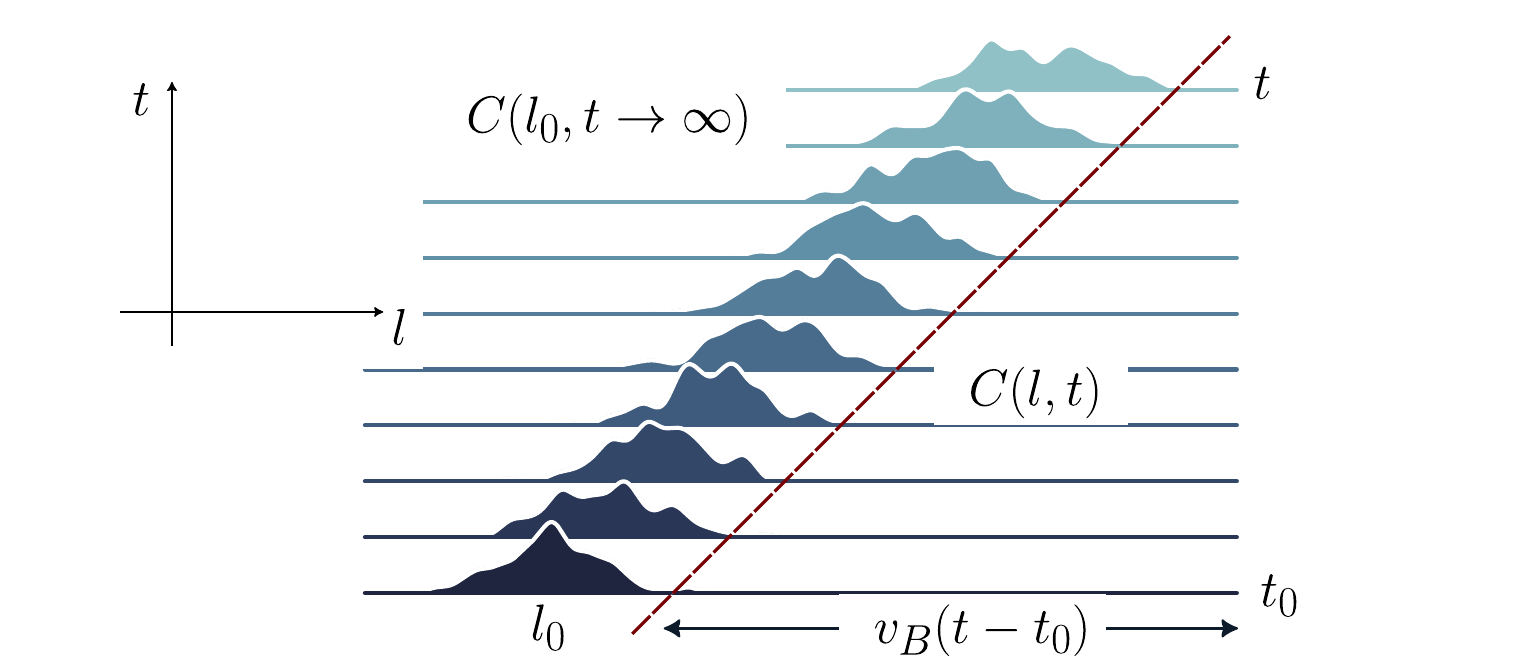}
	\caption{Schematic showing the ``shell-like'' structure of OTOC in which as the wavefront passes the operator site $l$, the OTOC approaches to zero in the long-time limit. The horizonal line shows different time slices increasing in an upward direction. We note that the site where OTOC is non-zero at initial time, goes to zero at late time, implying that the operator has no support on site.}
	\label{fig:illu}
\end{figure}

In this article, we study the out-of-time-ordered correlator (OTOC) and K-complexity for different operators in the exactly solvable one-dimensional discrete-time quantum walk. Our study of OTOC for different spin operators shows a universal ``shell-like'' structure so as the wavefront passes, the OTOC goes to zero in the long-time limit, in other words, the operator has no support on the site, implying the absence of  scrambling (see Fig.~\ref{fig:illu}). On the other hand, we show that K-complexity grows linearly in time, akin to the approximate orthonorgonality of operator at each time-step. We further study the effect of disorder which generally results in the slowdown of information scrambling. In both spatial and temporal disorder, the shape of the light-cone deforms showing no scrambling beyond localization length. The K-complexity growth transititions from linear to sub-linear showing saturation at late times, thus reflecting the localization of the operator.

\section{Quantifying scrambling}
\subsection{Out-of-time-ordered correlators}
OTOCs provide a means to quantify the evolution of operators. Let's consider two local operators, $W$ and $V$, within a one-dimensional spin chain. The idea is to probe the spread of $W(t) = e^{iHt} W e^{-iHt}$ using another operator $V$, typically a simple spin operator positioned at a distance $l$ from $W$ evolving under system Hamiltonian $H$. To do this, we considers the expectation value of the squared commutator,
\begin{equation}
	C(l,t) = \left \langle [W(t),V]^\dagger [W(t),V]\right\rangle
\end{equation}
Initially, this quantity is zero for widely separated operators, but it deviates significantly from zero once $W(t)$ extends to the location of $V$. In the special scenario where $W$ and $V$ are Hermitian and unitary, the squared commutator can be expressed as $C(t) = 2- 2\text{ Re } [\langle W(t) VW(t) V\rangle]$, with $\langle W(t) V W(t) V\rangle = F(t)$ representing the OTOC. The growing interest in OTOCs has spurred numerous experimental proposals and experiments aimed at measuring them\,\cite{PhysRevA.94.040302,doi:10.1126/science.abg5029,PhysRevLett.129.160602,garttner_measuring_2017,braumuller_probing_2022,landsman_verified_2019}.

The OTOC serves as a tool to investigate characteristics of chaos such as operator growth and the butterfly effect. In the case of local interactions, the growth $C(l,t)$ is conjecture to obey\,\cite{xu_accessing_2020,PhysRevX.9.031048,PhysRevB.98.144304,PhysRevLett.123.165902}
\begin{equation}
	C(l,t) \sim \exp \left[-\lambda_p \frac{\left(l-v_Bt\right)^{1+p}}{t^p} +a \log t\right]
\end{equation}
Here, $v_B$ denotes the butterfly velocity, $p$ represents the wavefront broadening coefficient, and $a$ encapsulates the logarithmic growth observed in the many-body localized (MBL) phase under disorder.

\subsection{Krylov complexity}\label{sec:Krylov complexity}
\subsubsection{Continuous-time evolution}\label{subsubsec:continuous-time evolution}

In a closed system, the evolution of any operator $\mathcal{O}_0$ under a time-independent Hamiltonian $H$ is described by the Heisenberg equation of motion,
\begin{equation}\label{eq:operator_evolution}
	\mathcal{O}(t) = e^{itH}\mathcal{O}_0 e^{-itH} = e^{i\mathcal{L}t} O_0 = \sum_{n=0}^\infty \frac{(it)^n}{n!}\mathcal{L}^n \mathcal{O}_0
\end{equation}
where $\mathcal{L}$ is the Hermitian Liouvillian superoperator given by $\mathcal{L} = [H,\,\bullet\, ]$. Therefore, the operator $\mathcal{O}(t)$ can be written as a span of the nested commutators with the initial operator \textit{i.e.} $\{\mathcal{L}^n \mathcal{O}_0\}_{n=0}^\infty$. An orthonormal basis $\{|O_n)\}_{n=0}^{K-1}$ can be constructed from this nested span of commutators, by choosing a certain scalar product $(\cdot|\cdot)$ on operator space using a form of Gram-Schmidt orthogonalization known as the Lanczos algorithm. The dimension of Krylov space $K$ obeys a bound $K \leq D^2-D+1$, where $D$ is the dimension of the state Hilbert space\,\cite{rabinovici_operator_2021}. In the Krylov basis $\{|O_n)\}$, the Liouvillian takes the tridiagonal form  $\mathcal{L}|O_n) = b_{n+1}|O_{n+1}) + b_n |O_n)$, where $b_n$ are known as the Lanczos coefficients that are tied to chaotic nature of the system at hand\,\cite{rabinoviciKrylovLocalizationSuppression2022}. We can write the expansion of the operator $\mathcal{O}(t)$ in terms of the constructed Krylov basis as
\begin{equation}
	\mathcal{O}(t) = \sum_{n = 0}^{K-1} i^n \phi_n(t) |O_n)
\end{equation}
The amplitudes $\phi_n(t)$ evolve according to the recursion relation $\dot{\phi}_n(t) = b_{n-1}\phi_{n-1}(t) - b_n\phi_{n+1}(t)$ with the initial conditions $\phi_n(0)=\delta_{n,0}$\,. The recursion relation suggests that the Lanczos coefficients $b_n$ are hopping amplitudes for the initial operator $\mathcal{O}_0$ localized at the initial site to explore the \textit{Krylov chain}. With time, the operator gains support away from the origin in the Krylov chain reflects the growth of complexity as higher Krylov basis vectors are required in operator expansion. To quantify this, one defines the average position of the operator in the Krylov chain --- called the Krylov complexity as 
\begin{equation}\label{eq:complexity-continous}
	K(t) = (\mathcal{O}(t)|\mathcal{K}|\mathcal{O}(t)) = \sum_{n=0}^{K-1} n|\phi_n(t)|^2 
\end{equation}
where $\mathcal{K} = \sum_{n=0}^{K-1}n|O_n)(O_n|$ is the position operator in the Krylov chain. For our purpose, we will use the infinite-temperature inner product, also known as the Frobenius inner product :
\begin{equation}
	(\mathcal{A}|\mathcal{B}) = \frac{1}{D} \text{Tr}\left[\mathcal{A}^\dagger \mathcal{B}\right],\ \ \ \lVert \mathcal{A} \rVert = \sqrt{(\mathcal{A}|\mathcal{A})}\,.
\end{equation} 

\subsubsection{Discrete-time evolution}
To formulate the K-complexity for a system with discrete-time evolution such that $\mathcal{O}_n  = U^\dagger_n \mathcal{O}_{n-1} U_n = \mathcal{U}_n[\mathcal{O}_n]$, where $n = 1,2,\ldots$, $U_n$ is the unitary operator which describes the evolution of the system at time-step $t = nT$ with step-size $T$, and $\mathcal{U}_n$ is the unitary superoperator given by $\mathcal{U}_n = U_n^\dagger \bullet U_n$. We define the Krylov basis by choosing $|O_0) = \mathcal{O}_0$ and then recursively orthogonalizing each $\mathcal{O}_n$ with all the $|O_n)$ for $i < n$. This choice of basis (represented by $\mathbb{O}$) maximizes the cost function defined as\,\cite{PhysRevD.106.046007}
\begin{equation}
	K_\mathbb{B}(t) = \sum_i \zeta_i \lvert ( \mathcal{O}_t | B_i) \rvert^2
\end{equation}
with respect to an arbitrary choice of orthonormal set $\mathbb{B} = \{|B_i): i = 0,1,2,\ldots\}$. The coefficients $\zeta_i$ (referred to as the ``weight function'') are positive, increasing the sequence of real numbers. The optimal choice of the basis follows from the induction method. We fix the initial state as the first state of the Krylov basis $\mathcal{O}_0 = |O_0) $. Assume the first $N$ vectors of certain basis $\mathbb{B}$ are same as the Krylov basis, i.e., $|B_i\rangle = |O_i\rangle $ for $i = 0,1,\ldots ,N-1$. By assumption 
$$\mathcal{O}_n = \sum_{j = 0}^{N-1}( K_j |\mathcal{O}_n) |O_j)\qquad n\leq N-1\,.$$
Therefore, the cost associated with the two basis are the same for $n\leq N-1$. The next state $|O_N\rangle $ into a part belonging to the Krylov subspace $|O_i\rangle $, for $i \leq N-1$, and a part perpendicular to it \textit{i.e.}
$$\mathcal{O}_N = p_\perp |O_N) + p_\parallel |O_\parallel), \qquad |O_\parallel\rangle = \sum_{i=0}^{N-1} a_i |O_i)$$
where $|O_N)$ is the next element of the Krylov basis by definition. A basis different from the Krylov one would necessarily not include $|O_N\rangle$. Therefore, the cost at discrete time $N$ would be larger, since we would have to express $|K_N\rangle$ in the new basis, which would require at least two vectors. Since the contribution to the cost from the part $|O_\parallel\rangle $ is the same in both bases, the cost must increase when we divide $|O_N\rangle$ into several contributions, since $\zeta_n$ is a strictly increasing function of $n$. Therefore, the Krylov basis minimizes the cost function for all times.\\

\noindent At any time $t=nT$, we can expand the state $\mathcal{O}_t$ in the Krylov basis as
\begin{equation}
	\mathcal{O}_n = \sum_{i=0}^{D^2} \phi_{i,n} |O_i)\,,
\end{equation}
where the expansion coefficient $\phi_{i,t} = (O_i|\mathcal{O}_n)$. We define the K-complexity of the state as the average position of the distribution on the ordered Krylov basis:
\begin{equation}\label{eq:complexity-discrete}
K(t) = \sum_{i=0}^{D^2} i|\phi_{i,n}|^2 \,.
\end{equation}
Since the evolution operator itself is used as a generator of the Krylov basis, we can bound the maximum possible growth of the K-complexity. \textit{For any system evolving unitarily in discrete-time, the K-complexity can grow at most linearly with time $t$.} To prove this, we consider maximizing the complexity at anytime $t$ with respect to the expansion coefficients, i.e.,
\begin{equation}
\max_{\{\lvert \phi_{i,n}\rvert \}} K(t)  = \max_{\{\lvert \phi_{i,n}\rvert \}} \sum_{i}  i|\phi_{i,n}|^2 
\end{equation}
with constraint $\sum_i |\phi_{i,n}|^2 = 1$. From the orthonormalization construction, it follows that the expansion coefficients $\phi_{i,n} = 0 \ \forall \ i>n$. Therefore, it follows that the complexity can at most grows linearly with time $t$ corresponding to the case where $\phi_{i,n} = \delta_{i,n}$.  In terms of the operator, it means that $(O_{n'}|O_n)\ \forall \ n'<n$ which corresponds to the maximally ergodic regime universal in chaotic systems\,\cite{Suchsland:2023cmb}.      

In the case where the evolution operator is time-independent, we can define a quasi-Hamiltonian $H_q$ such that $U = e^{-i H_q T}$. In this case, the unitary operator $U$ takes upper Hessenberg form in the Krylov basis\,\cite{Suchsland:2023cmb}, given by 
\begin{equation}
(O_i|U|O_j) = \begin{cases}
0 & \text{if} \ j> i+1, \\
b_j & \text{if} \ j = i+1, \\
a_jc_i/c_j & \text{if} \ j<i+1,
\end{cases}
\end{equation}
with $a_0 = c_0$. The quasi-Hamiltonian $H_q$ can itself be used in the continuous-time setting to define complexity using the notion discussed in Sec.~\ref{subsubsec:continuous-time evolution}. To investigate the relation between the complexities, we consider the following expansion
\begin{equation}\label{eq:taylor-expansion}
e^{i\mathcal{L}_qT} = \sum_{k = 0}^\infty \frac{(iT)^k}{k!}\mathcal{L}_q^k \approx \sum_{k=0}^{K_c} \frac{(iT)^k}{k!}\mathcal{L}_q^k
\end{equation} 
where $\mathcal{L}_q$ is Liouvillian superoperator associated with the Hamiltonian $H_q$. The Taylor series is truncated at order $K_c$ which depends on the operator $\mathcal{L}_q$, step-size $T$ and error tolerance\,\cite{PhysRevLett.114.090502}. From Eq.\,\eqref{eq:taylor-expansion}, it follows that the operator at first time-step $\mathcal{O}_1$ will be linear combination of operators $\{\mathcal{L}^k_q \mathcal{O}_0\ | \ k = 0,1,\ldots ,K_c\}$. In other words, the first Krylov basis vector
\begin{equation}\label{eq:operator-relation}
|O_1) = \sum_{k = 0}^{K_c} \psi_{k,1} |O^{(q)}_k) 
\end{equation}
where the set $\mathbb{O}^{(q)} = \{|O^{(q)}_k)\ |\ k = 0,1,\ldots \}$ corresponds to the Krylov basis associated with the Liouvillian $\mathcal{L}_q$. In general, we will assume that there exists a cutoff value $K_c^{(i)}$ such that 
\begin{align}
|O_i) &= \sum_{k = 0}^{K_c^{(i)}} \psi_{k,i} |O_k^{(q)}) \\
\mathcal{O}_n &= \sum_{i = 0}^n \phi_{i,n} |O_i)  = \sum_{k = 0}^{K_c^{(n)}} \phi^{(q)}_{k,n} |O^{(q)}_k)\,.
\end{align}
Therefore, the K-complexity associated with the Liovillian $\mathcal{L}_q$ (refer to as \textit{quasi K-complexity}) given by
\begin{equation}
\begin{split}
K^{(q)}(t) &= \sum_{k = 0}^{K_c^{(n)}} k |\phi_{k,n}^{(q)}|^2  = \sum_{k = 0}^{K_c^{(n)}} k \left|\sum_{i = 0}^n \phi_{i,n}\psi_{k,i}\right|^2 \\
&\leq  \sum_{k = 0}^{K_c^{(n)}} k \sum_{i=0}^n |\phi_{i,n}|^2|\psi_{k,i}|^2\\
&= \sum_{i=0}^n \left(\sum_{k = 0}^{K_c^{(n)}} k|\psi_{k,i}|^2 \right) |\phi_{i,n}|^2\,.
\end{split}
\end{equation}
Therefore, the quasi K-complexity $K^{(q)}(t)$ is similar to that of the unitary operator with a modified weight factor $\zeta_i$. The suitable choice of weight factor to recover the quasi K-complexity is non-trivial and depends on growth of wave functions $\phi^{(q)}_{k,n}$. Let us consider case where the wave function grows exponentially $|\phi^{(q)}_{k,t}|^2 \sim  e^{-k/\xi(t)}$, where $\xi(t)$ is delocalization length that grows exponentially in time $\xi(t)\sim e^{2\alpha t}$ for $\alpha t \gg 1$, corresponding to exponential growth (maximum possible growth) of quasi K-complexity\,\cite{PhysRevX.9.041017} $K^{(q)}(t)\sim e^{2\alpha t}$. To retain this growth, one possible choice of weight function is $\zeta_i = e^{2\alpha i}$ along with maximal growth of wave function $\phi_{i,n} = \delta_{i,n}$.\\

We will now consider the limiting case where step-size $T \rightarrow 0$ to recover the continuous limit. In this case, $e^{i\mathcal{L}_qT} \approx I + iT \mathcal{L}_q$, up to correction of order $\mathcal{O}(T^2)$, therefore, the krylov basis generated by unitary operator $U$ and quasi-Liouvillian $\mathcal{L}_q$ matches exactly to each other. However, it should be noted that the two complexities will eventually start to differ at large time-step as error start to accumulate and grow large. 

\section{Discrete-time quantum walks} The discrete-time quantum walk on a line is defined on a Hilbert space $\mathcal{H} = \mathcal{H}_c\otimes \mathcal{H}_p$ where $\mathcal{H}_c$ is coin Hilbert space and $\mathcal{H}_p$ is the position Hilbert space. For a walk in one dimension, $\mathcal{H}_c$ is spanned by the basis set $\lvert\uparrow\rangle $ and $\lvert\downarrow\rangle$ representing the internal degree of the walker, and $\mathcal{H}_p$ is spanned by the basis state of the position $|x\rangle $ where $x\in \mathbb{Z}$ on which the walker evolves. At any time $t$, the state can be represented by 
\begin{equation}
	|\Psi(t)\rangle = \lvert \uparrow \rangle \otimes |\Psi^\uparrow(t)\rangle  + \lvert \downarrow\rangle \otimes | \Psi^\downarrow(t)\rangle  \\
	= \sum_x \begin{bmatrix}
		\psi^\uparrow_{x,t} \\
		\psi^\downarrow_{x,t}
	\end{bmatrix}.
\end{equation}
Each step of the discrete-time quantum walk is defined by a unitary quantum coin operation $C$ on the internal degrees of freedom of the walker followed by a conditional position shift operation $S$ which acts on the configuration of the walker and position space. Therefore, the state at time $(t+1)$ will be
\begin{equation}
	|\Psi(t+1)\rangle = S(C\otimes I) |\Psi(t)\rangle= W|\Psi(t)\rangle.
\end{equation}
The general form of coin operator $C$, given by 
\begin{equation}
	C = C(\xi,\theta, \varphi ,\delta ) = e^{i\xi}e^{-i\theta \sigma_x}e^{-i\varphi \sigma_y}e^{-i\delta \sigma_z} 
\end{equation}
where $\xi$ is global phase angle, $2\theta,2\varphi, 2\delta$ are the angles of rotations along $x,y$ and $z$ axes respectively with $\theta,\varphi,\delta \in [0,2\pi]$, and $\sigma^\mu$ is the $\mu^{\text{th}}$ component of the Pauli spin matrices $\{\sigma_x,\sigma_y,\sigma_z\}$, which are generators of SU(2) group. The position shift operator $S$  is of the form
\begin{equation}
	S = \lvert \downarrow\rangle \langle \downarrow\rvert  \otimes T_+ + \lvert\uparrow\rangle \langle \uparrow \rvert \otimes T_-,  \quad T_\pm = \sum_{x \in \mathbb{Z}} |x\pm 1\rangle \langle x|
\end{equation}
are translation operators. In this work, we will consider the specific choice of coin operator
\begin{equation}
	C(\theta) = \begin{bmatrix}
		\cos \theta & \sin \theta \\
		-\sin \theta & \cos \theta 
	\end{bmatrix} 
\end{equation}
corresponding to parameters  $(0,\theta,0,3\pi/2)$ which previously was studied in the context of the Dirac dynamic\,\cite{chandrashekarTwocomponentDiraclikeHamiltonian2013}. In momentum basis, the unitary operator can be diagonalized to obtain the dispersion relation in space-time continuum limit\,\cite{chandrashekar2013disorder}
\begin{equation}
	\omega(k,\theta) = \pm \sqrt{k^2 \cos \theta + 2(1-\cos \theta)}
\end{equation}
Therefore, the group velocity $v_g(k,\theta)$ given by 
\begin{equation}
	\begin{split}
		v_g(k,\theta) &\equiv \frac{d\omega(k,\theta)}{dk} \\
		&= \pm \frac{k\cos \theta}{\sqrt{k^2 \cos\theta + 2(1-\cos \theta)}}\,.
	\end{split}
\end{equation}
It's important to note that group velocity is maximum (equals to $1$) at $\theta = 0$ for all $k$ which corresponds to identity as coin operator. The coin parameter $\theta$ controls the variance $\sigma^2$ of the probability distribution in the position space and this distribution spreads quadratically faster ($\sigma^2 \approx  [1 - t^2\sin\theta ])$ in position space when compared to the classical random walk\,\cite{PhysRevA.77.032326}.

\subsection{Disordered discrete-time quantum walk}

There are number of ways to induce disorder in discrete-time quantum walk that usually lead to localization of wavefunction\,\cite{PhysRevA.83.022320,PhysRevB.101.064202,PhysRevB.96.144204,PhysRevB.101.144201,zeng_discrete-time_2017,kumar_enhanced_2018}. Here, we will consider two choices of disorder --- spatial  and temporal disorder. The spatial disorder in quantum walk is define by introducing a position dependent coin operator $C(\theta_x)$ with $\theta_x \in \theta_0 + \{-W/2,W/2\}$ where $0\leq W\leq \pi$ defines the disorder strength and $\theta_0$ is mean value. Therefore, the evolution of the state  is described by
\begin{equation}
	|\Psi(t+1)\rangle  = S\cdot \bigoplus_x C(\theta_x) |\Psi(t)\rangle \,.
\end{equation}
In the similar analogy, the temporal disorder in quantum walk define by introducing a time-dependent coin operator $C(\theta_t)$ with $\theta_t \in \theta_0 + \{-W/2,W/2\}$. The evolution of the state is given by 
\begin{equation}
	|\Psi(t+1)\rangle  = S\cdot (C(\theta_t)\otimes I) |\Psi(t) \rangle\,.
\end{equation}
While the temporal disorder in quantum walk leads to a weak localization, the spatial disorder is known to induce Anderson localization\,\cite{PhysRevB.96.144204}. Although, in both cases,
\begin{equation}
	\lim_{t\rightarrow\infty} \langle v_g^{\text{SD/TD}} \rangle \rightarrow 0
\end{equation}
The mean group velocity drops to zero faster for a walk with spatial disorder resulting in strong localization compared to temporal disorder which leads to weak localization. the localization length is usually a function of the coin parameter $\theta$ given as $\zeta = [\ln \cos \theta]^{-1}$. Both of these disorders were studied extensively in enhancing the entanglement and non-Markovianity generated between the internal and external degrees of freedom\,\cite{PhysRevLett.111.180503,kumar_enhanced_2018}. 

\section{Results}

\subsection{Out-of-time-ordered correlator}\label{subsec:OTOC}
We will be interested in the quantities 
\begin{equation}\label{eq:OTOC}
	\begin{split}
		C_{\mu\nu}(l,t) &\equiv \frac{1}{2} \left\langle \left| [W_l^\mu(t), V^\nu_0] \right|^2 \right\rangle \\
		& = \frac{1}{2} \left \langle [W^\mu_l(t),V^\nu_0]^\dagger [W^\mu_l(t),V^\nu_0]\right \rangle 
	\end{split}
\end{equation}
where $\mu,\nu \in \{x,y,z\}$ and the operators $W^\mu_l$ and $V^\nu_0$ are local operators defined as $\sigma^\mu \otimes |l\rangle \langle l|$ and $\sigma^\nu \otimes |0\rangle \langle 0|$, respectively.
\begin{figure*}
	\centering
	\includegraphics[width=0.328\linewidth]{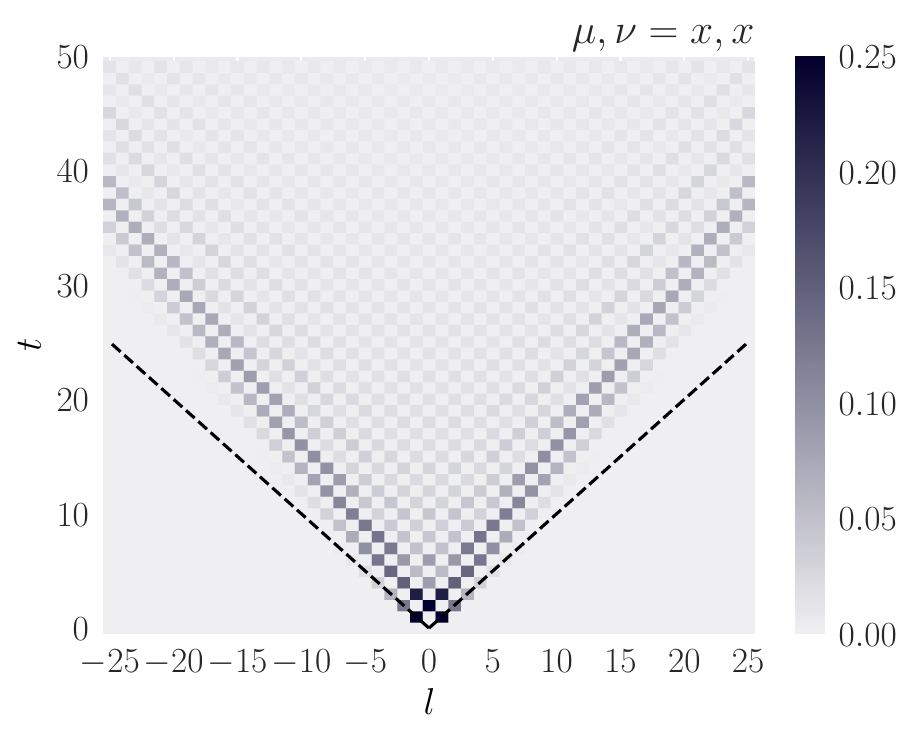}
	\includegraphics[width=0.328\linewidth]{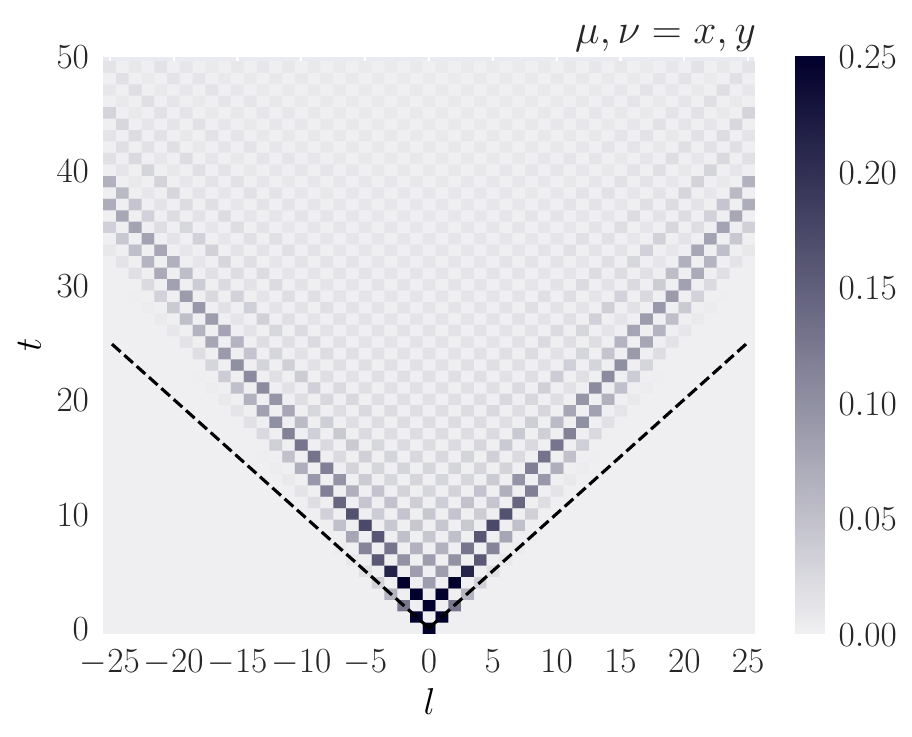}
	\includegraphics[width=0.328\linewidth]{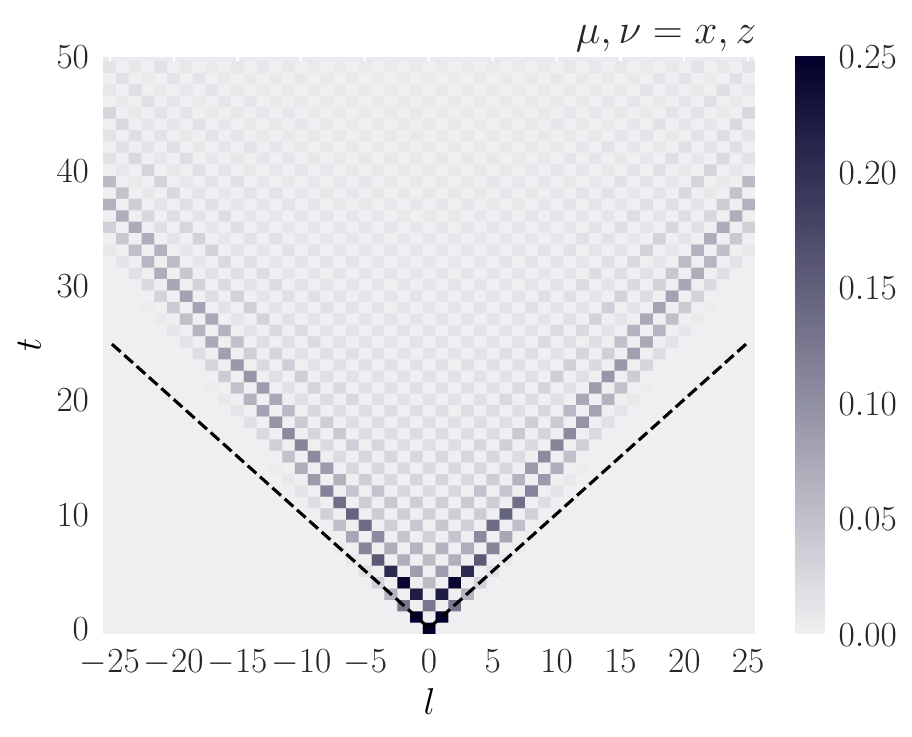}	
	\includegraphics[width=0.328\linewidth]{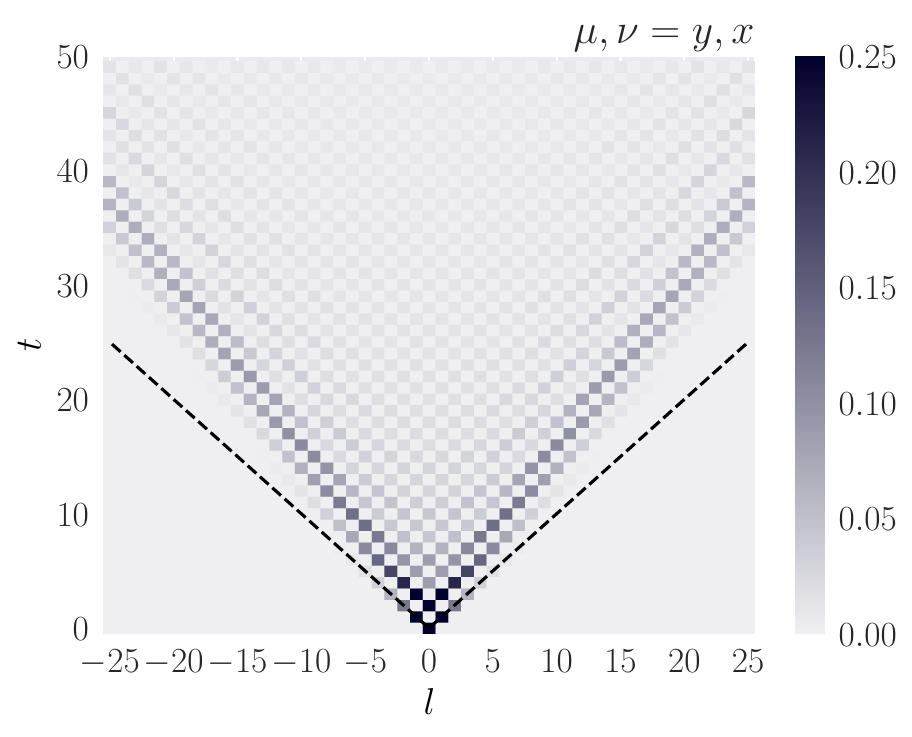}
	\includegraphics[width=0.328\linewidth]{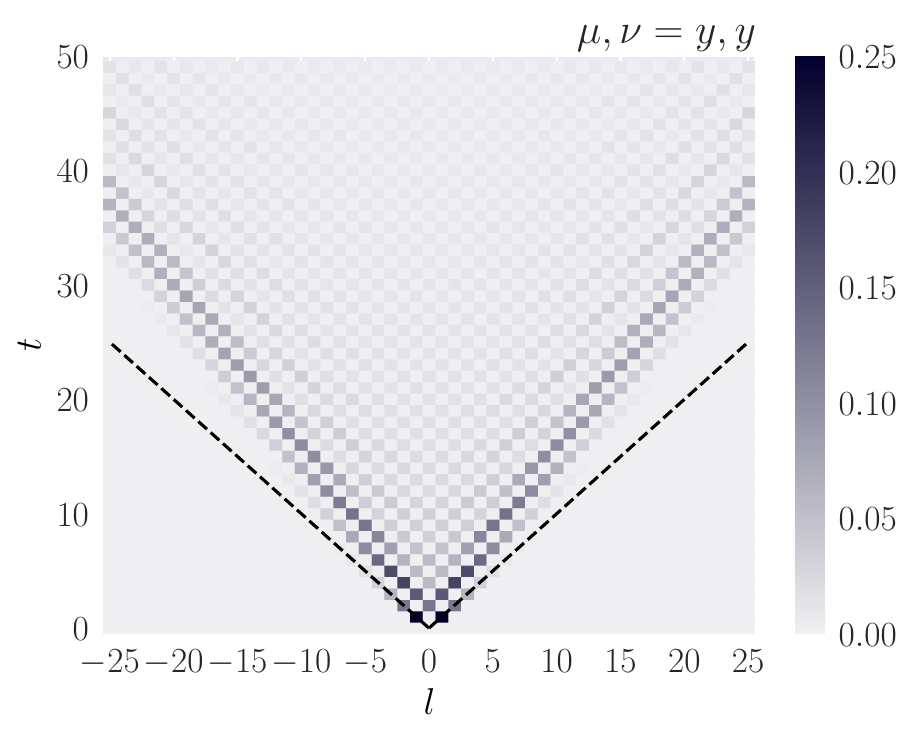}
	\includegraphics[width=0.328\linewidth]{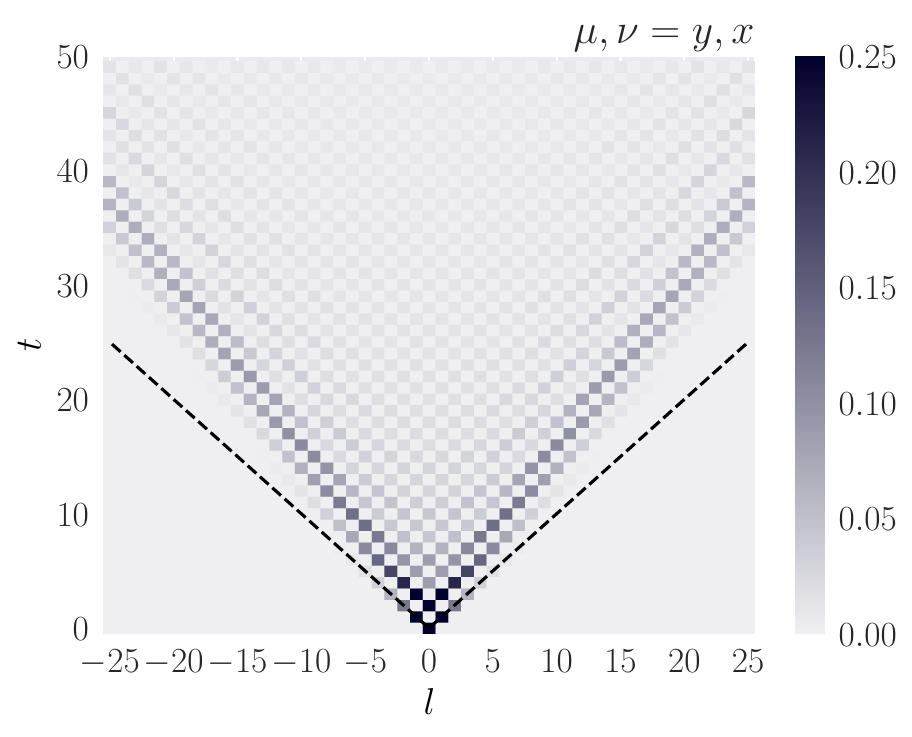}	
	\includegraphics[width=0.328\linewidth]{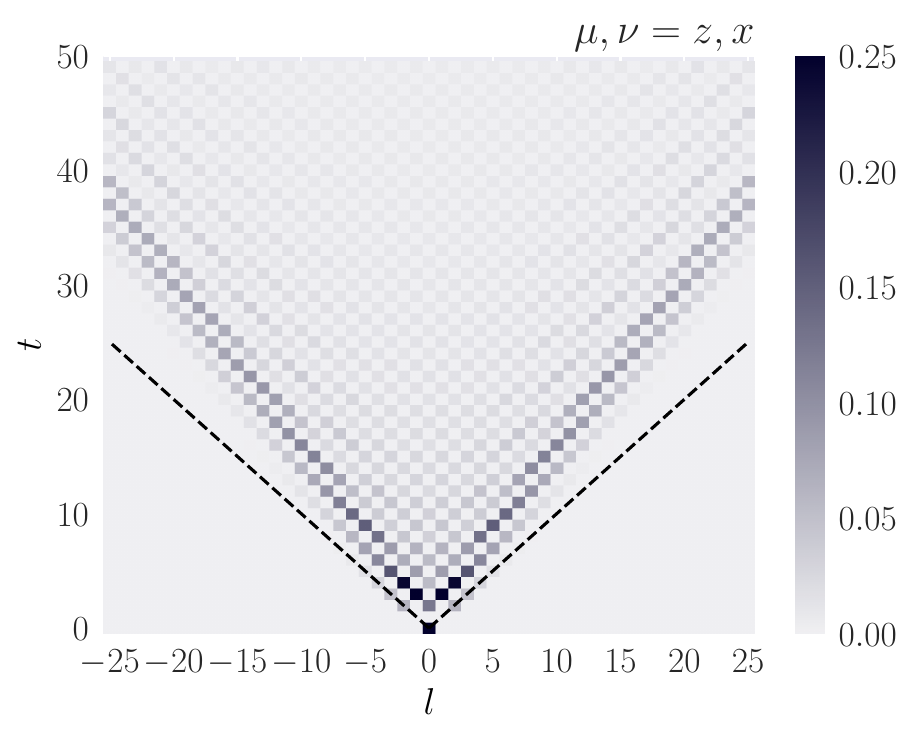}
	\includegraphics[width=0.328\linewidth]{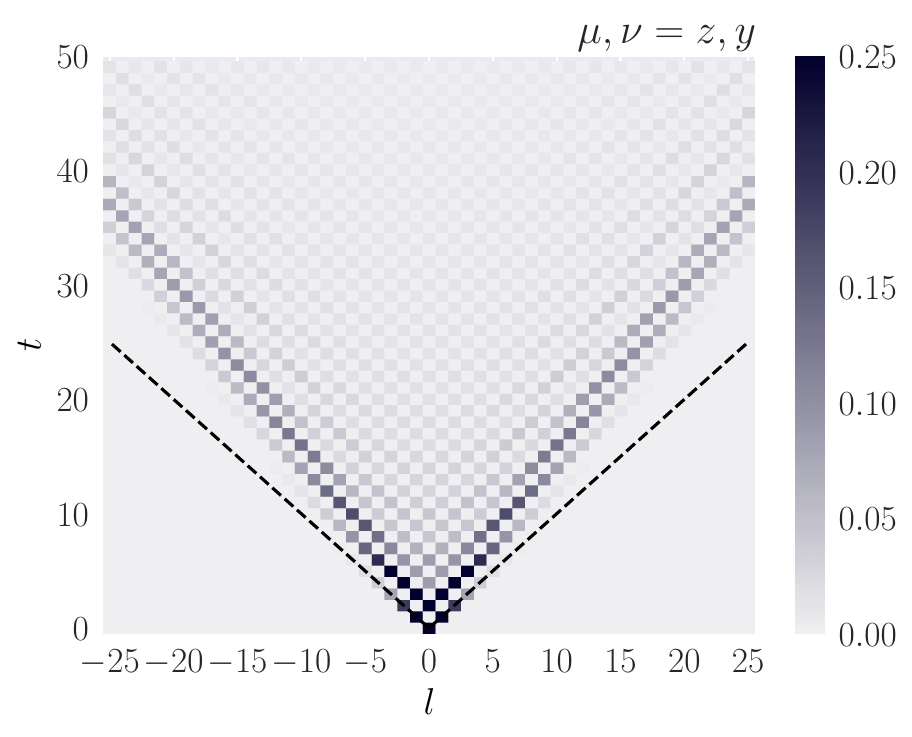}
	\includegraphics[width=0.328\linewidth]{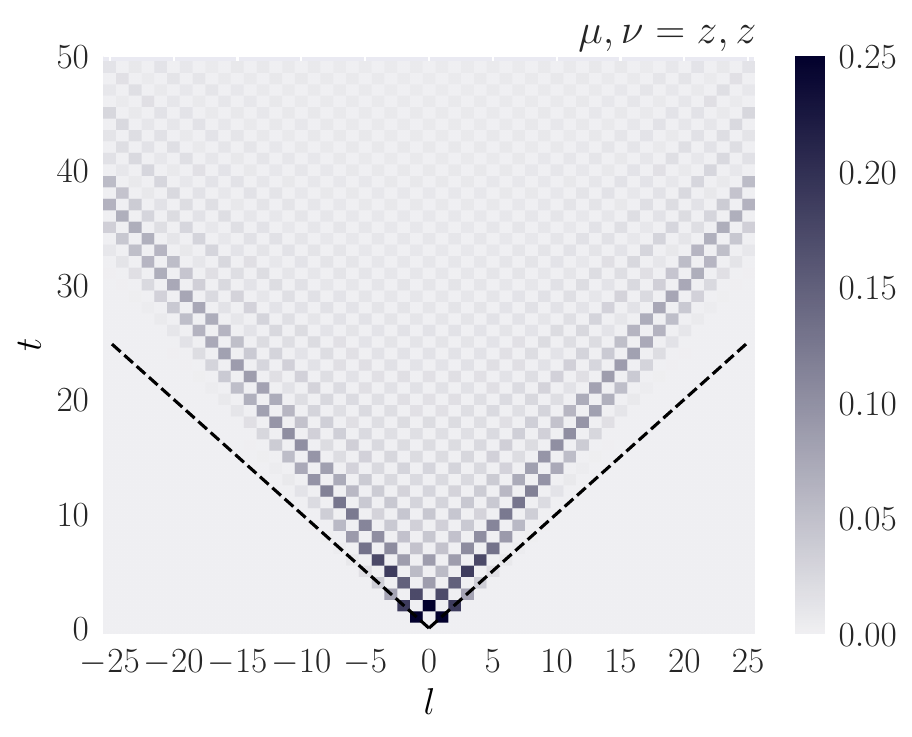}	
	\caption{The function $C_{\mu\nu}(l,t)$ for the discrete-time quantum walk; the system size is $L = 100$; the coin-angle $\theta = \pi/4$. We show data as matrix plot as a function of $l$ and time $t$ (the maximum is set to $0.25$ in all cases for better visibility and comparison). The light cone can be readily identified and corresponds to the maximal quasiparticle group velocity $v_B = \text{max}_k\ d\epsilon_k/dk$. In the timelike region, $C_{\mu\nu}(l,t)$ approaches zero in the long-time limit, indicating the absence of ``scrambling''. The dotted lines shows the light cone corresponding to case $v_B = 1$ corresponding to the case $\theta  = 0$}.
	\label{fig:otoc}
\end{figure*}

We consider the function $C_{\mu\nu}(l,t)$ in Eq.~\eqref{eq:OTOC} for the discrete-time quantum walk in one-dimension and coin-angle $\theta$ for varying distance $l$ between the initial operators. Figure~\ref{fig:otoc} shows the numerical results for $C_{\mu\nu}(l,t)$ at various time slices. We can identify the velocity of the wavefront as $v_B = \text{max}_k\ d\epsilon_k/dk = \text{max}_k\  v_g(k,\theta)$. It follows that light cone grows linearly with maximum velocity $v_B = 1$ corresponding to $\theta = 0$. In continuous systems, the long range interaction can result in super-linear growth of light-cone\,\cite{PRXQuantum.5.010201}. Therefore, in principle, the long range connectivity in discrete-time evolution can also result in super-linear growth of light-cone. In the present case, the OTOC function is ``shell-like''. That is, inside the timelike region, in the long-time limit, $C_{\mu\nu}(l,t)\rightarrow 0$, indicating no scrambling of operator $W^\mu_l(t)$, the vanishing of the $C_{\mu\nu}$ OTOC in the long-time limit suggests that expansion of $W^\mu_l(t)$ in terms of Pauli strings does not contain many pauli matrices ``in the middle'' of the strings. The feature common to Integrable quantum systems\,\cite{PhysRevB.97.144304}. Additionally, in case of quantum walk, we find that the functioon $C_{\mu\nu}$ oscillates between a finite value and zero which stems from the form of the shift operator and the choice of initial operators.

In integrable systems such as quantum spin systems\,\cite{PhysRevB.97.144304}, one can analytically find the squared commutator to exactly describe the nature of the wavefront and decay. In addition, although, quantum walk is also an integrable model, it turns out to be particularly hard to find out exact form of squared commutator. This is because the initial operators are local in position space, and the inverse fourier transform becomes hard due to the complexity of functions. However, in case of the Hadamard quantum walk ($\theta  = \pi/4$), we can use the known asymptotic results of wavefunction to vaguely argue the nature of decay and wavefront\,\cite{Nayak:2000xxw}. To this end, the asymptotics for the wave function can be described in three regions using $\alpha = l/t$. The wave function is essentially uniformly spread over the interval between $\mp 1/\sqrt{2} = \mp v_B$ where its gross behaviour is like $1/\sqrt{t}$. Outside this interval, the wave function dies out much faster than any inverse polynomial in $t$. At the ``wavefront'' ($\pm v_B$), there are two peaks of width $\mathcal{O}(t^{1/3})$, where the wave function goes as $t^{-1/3}$. If we approximately translate these wave function behaviour to operator $W^\mu_l(t)$, we conclude that the squared commutator decays as $\sim 1/t^2$ in the long-time limit and goes as $t^{-4/3}$ at the wavefront.

\begin{figure*}
	\centering
	\includegraphics[width=0.19\linewidth]{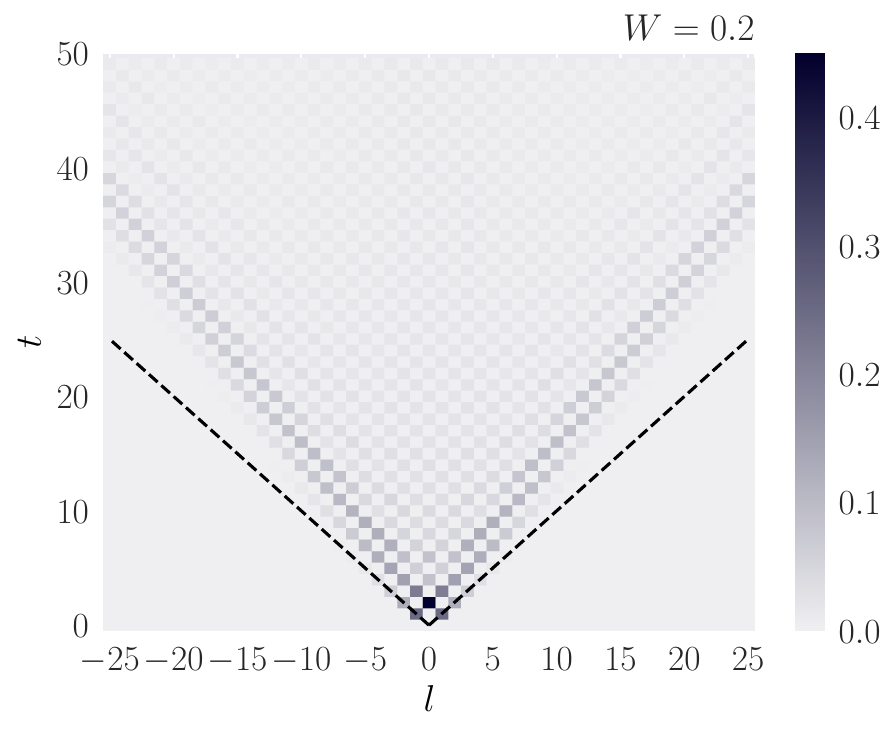}
	\includegraphics[width=0.19\linewidth]{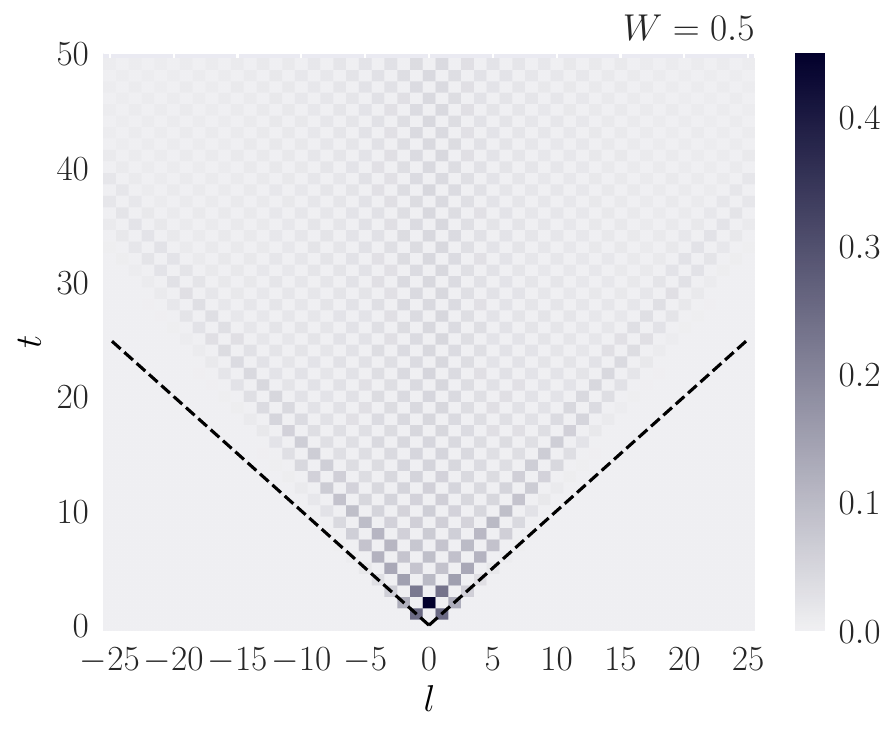}
	\includegraphics[width=0.19\linewidth]{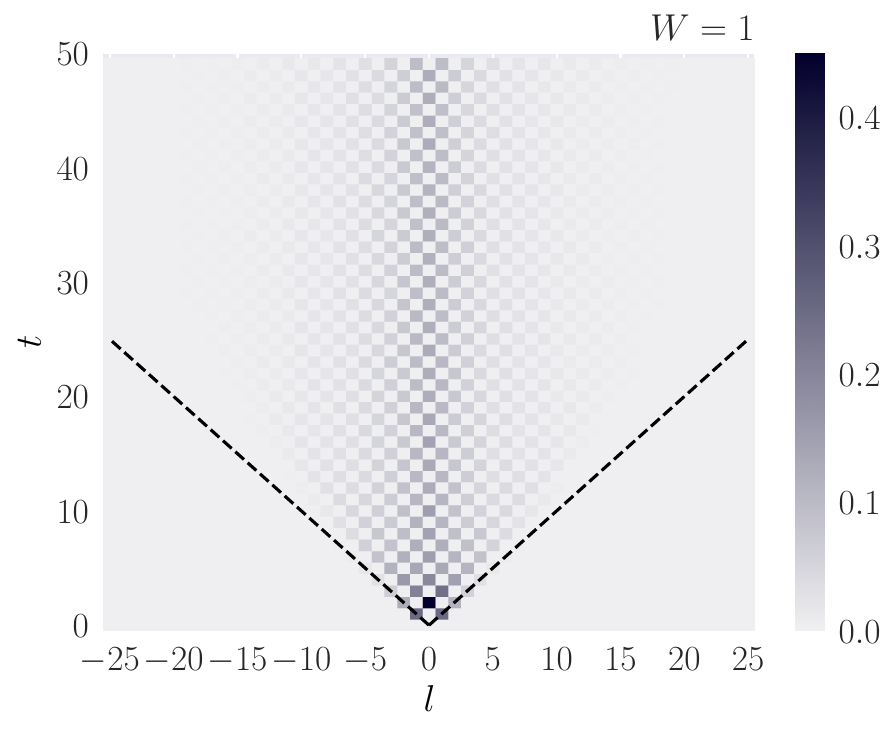}
	\includegraphics[width=0.19\linewidth]{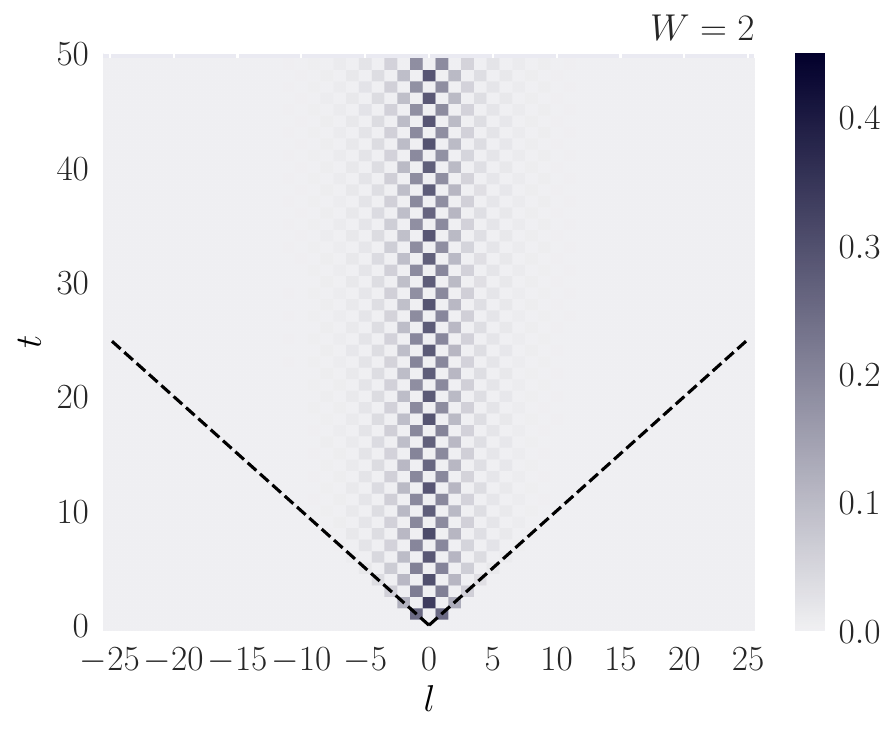}
	\includegraphics[width=0.19\linewidth]{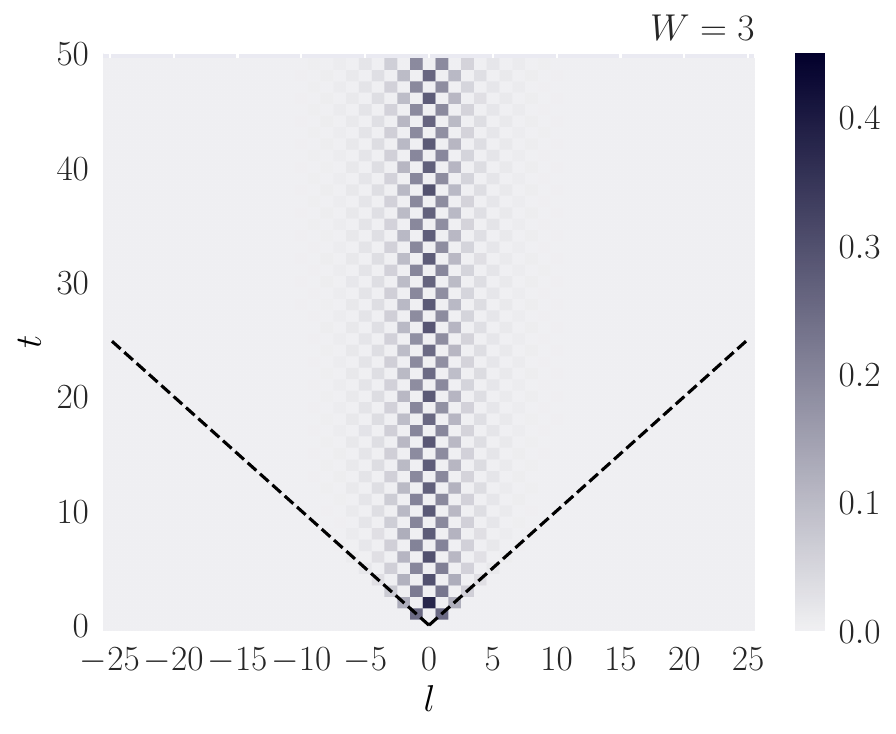}
	\includegraphics[width=0.19\linewidth]{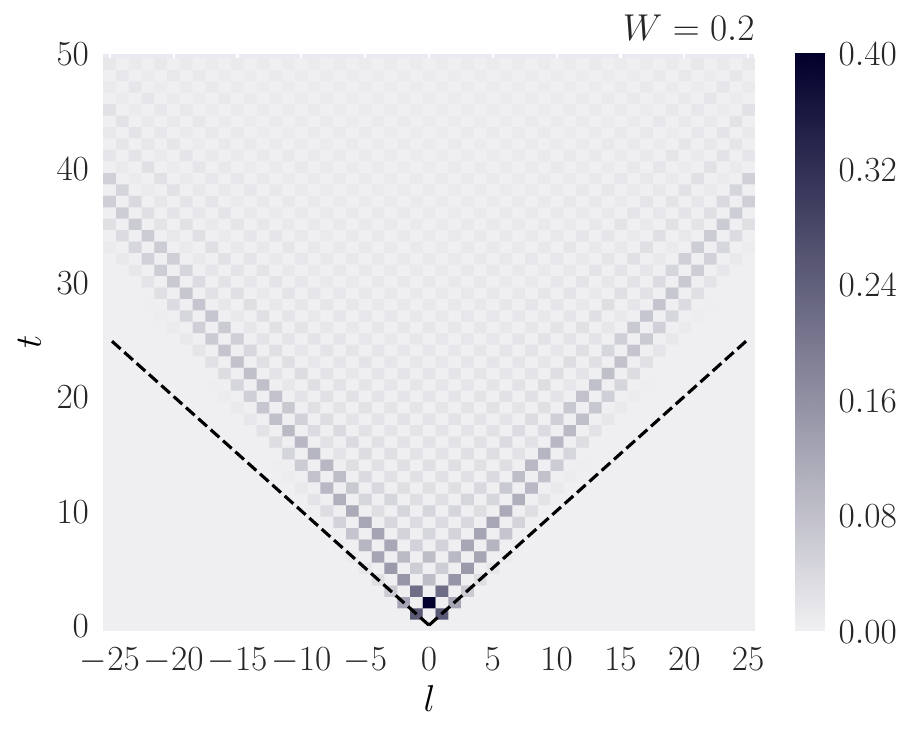}
	\includegraphics[width=0.19\linewidth]{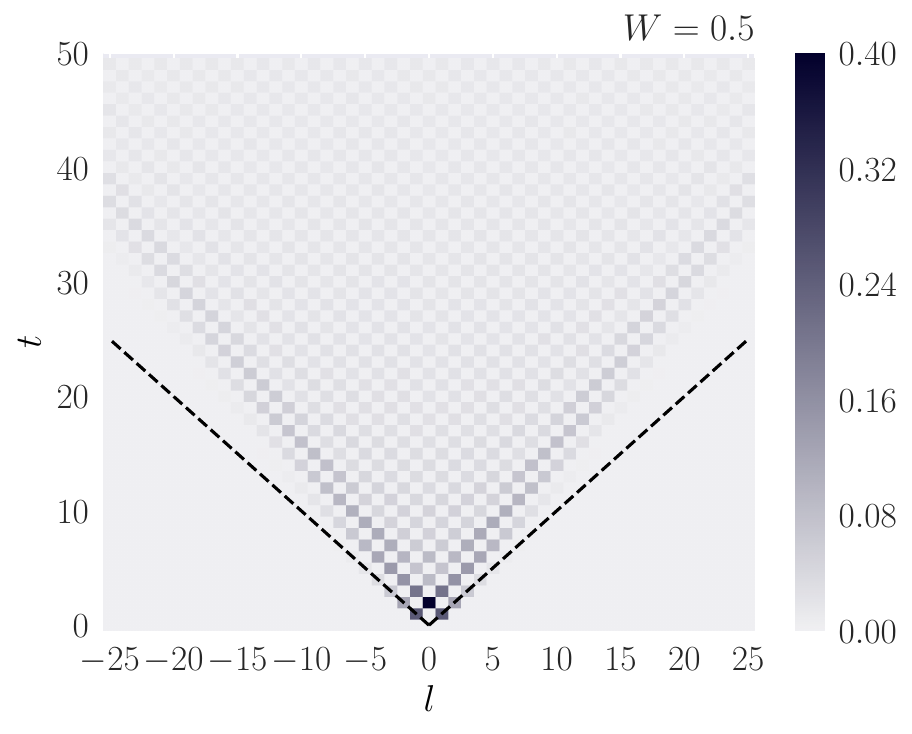}
	\includegraphics[width=0.19\linewidth]{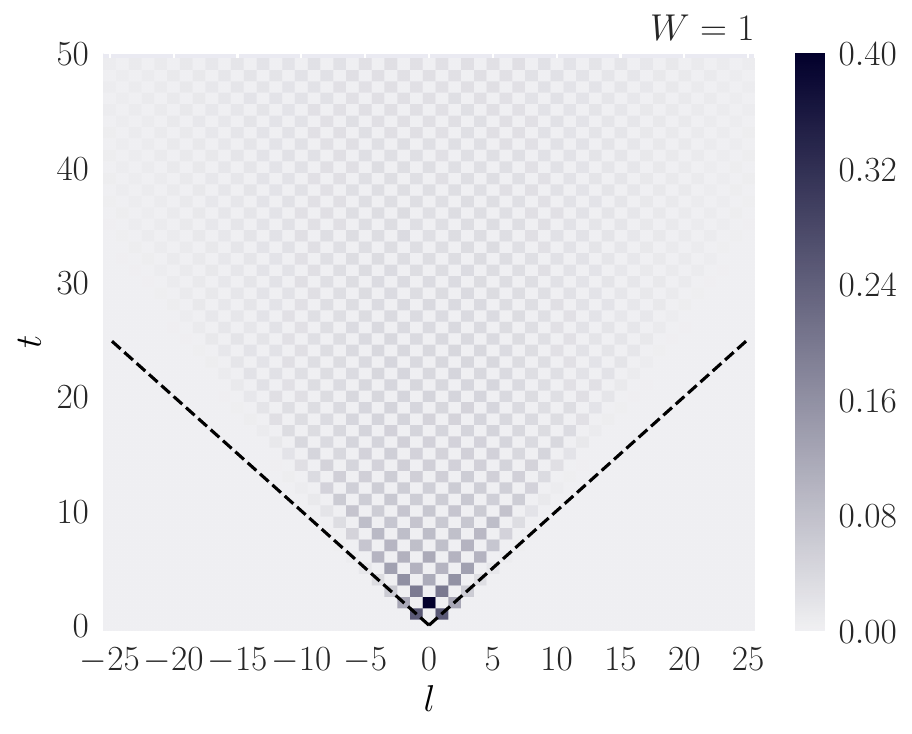}		
	\includegraphics[width=0.19\linewidth]{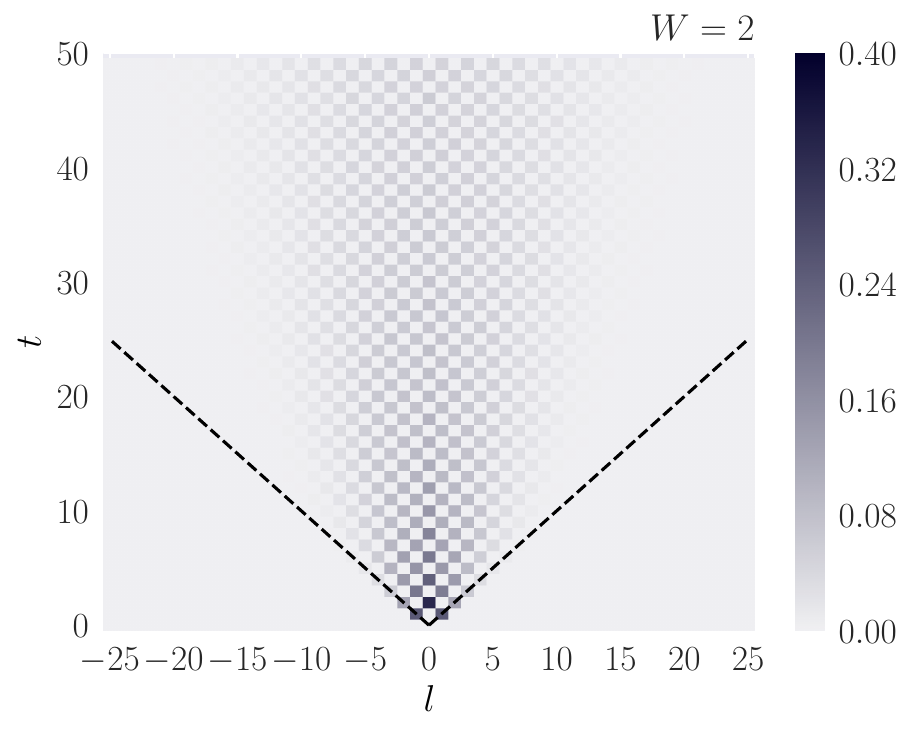}
	\includegraphics[width=0.19\linewidth]{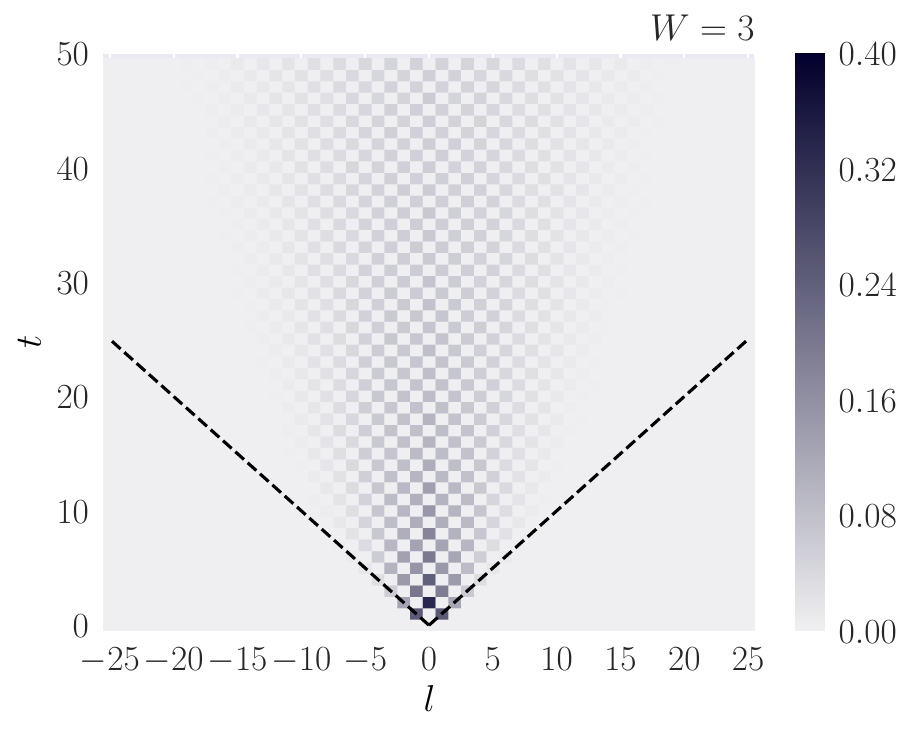}		
	\caption{The function $C_{xx}(l,t)$ for the discrete-time quantum walk in presence of spatial (top) and temporal (bottom) disorder; the system size is $L = 100$; the disorder strength $W$. The disorder average is taken over $500$ realizations.}
	\label{fig:OTOC_S}
\end{figure*}
The disorder (both spatial and temporal) in a system causes a slowdown in information propagation (see Fig.~\ref{fig:OTOC_S}). In particular, the spatial disorder results in Anderson localization which should be distinguished from MBL phase. The later is known as a non-interacting phenomenon. As the disorder strength increase, the shape of the light cone changes from ballistic to 
confine up to localization length for time $t \rightarrow \infty$ showing no information propagation beyond localization length. As shown in Fig.~\ref{fig:OTOC_S}, the spatial disorder leads to more rapid localization compared to temporal disorder as we increase disorder strength.    

\begin{figure}
	\centering
	\includegraphics[width=0.49\linewidth]{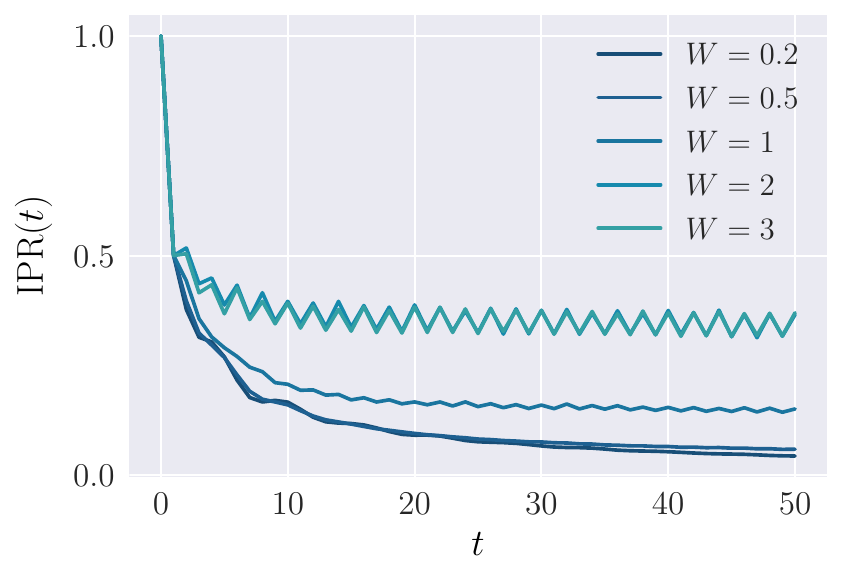}
	\includegraphics[width=0.49\linewidth]{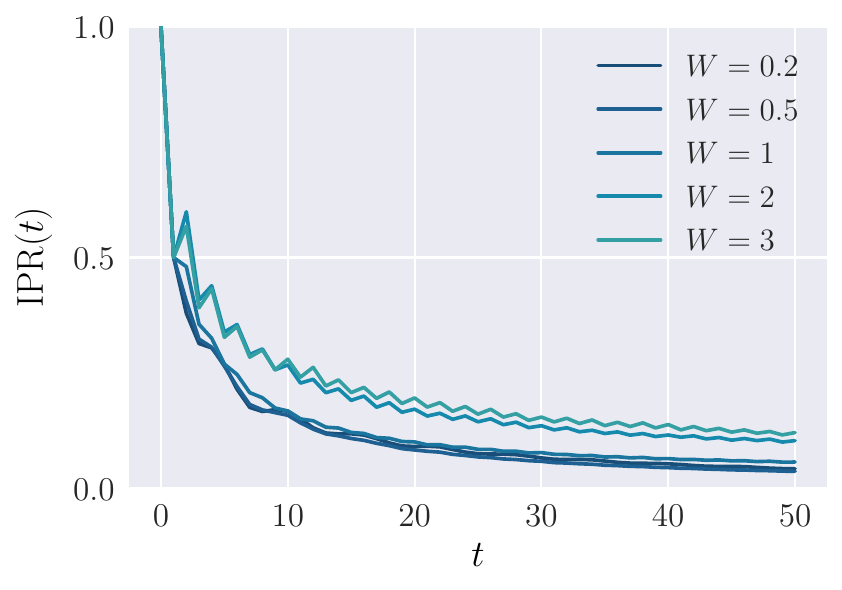}
	\caption{Inverse participation ratio calculated for spatial (left) and temporal (right) disorder with varying disorder strength $W$. The disorder average is taken over 500 realizations.}
	\label{fig:IPR}
\end{figure}

To characterize the localization, we consider inverse participation ratio (IPR) defined as\, \cite{RevModPhys.80.1355}
\begin{equation}
	\text{IPR}(t)  = \sum_x |\langle x|\Psi(t)\rangle |^4 = \sum_x p_x^2(t)
\end{equation}	
where $p_x(t)$ is probability of walker being at position $x$ and time $t$. The IPR quantifies the number of basis states that effectively contribute to the system's time evolution. Figure~\ref{fig:IPR} shows the IPR calculated in presence of spatial and temporal disorder for varying disorder strength $W$. In presence of disorder, the IPR saturates to a finite value which increase with disorder strength. The saturation value is larger in case of spatial compare to temporal showing that localization is strong in presence of spatial disorder. 

\subsection{K-complexity}	
Considering the formulation of K-complexity presented in Sec.~\ref{sec:Krylov complexity} for discrete-time evolution, we can show that the discrete-time quantum walk exhibit linear growth. We will consider the initial operator to be of the form $\mathcal{O}_0 = \sigma^\mu\otimes |0\rangle \langle 0|$. At any time $t$, the operator is given by (we will fix the step-size $T$ to unity.)
\begin{equation}
	\mathcal{O}_t = (U^\dagger)^t \mathcal{O}_0 U^t
\end{equation}
where the evolution operator $U = S(C\otimes I)$. We will show that the set of operators $\{\mathcal{O}_0,\mathcal{O}_1,\mathcal{O}_2,\ldots \}$ form a orthogonal basis i.e.  $( \mathcal{O}_i|\mathcal{O}_j) = \delta_{ij}$. First, we note that operators $\{ P_{ij} \equiv  |i\rangle \langle j| : i,j \in \mathbb{Z} \}$ forms a orthogonal basis in position Hilbert space $\mathcal{H}_p$. Next, consider the operator $\mathcal{O}_1 = U^\dagger \mathcal{O}_0 U$
\begin{align*}
	\mathcal{O}_1 &= (C^\dagger\otimes I \cdot S^\dagger) \mathcal{O}_0 (S \cdot C\otimes I) \\
	&= C^\dagger_\downarrow  \sigma^\mu C_\downarrow  \otimes |-1\rangle \langle -1| + C^\dagger_\downarrow \sigma^\mu C_\uparrow  \otimes |-1\rangle \langle 1|  \\
	& \qquad \qquad +  C^\dagger_\uparrow \sigma^\mu C_\downarrow  \otimes |1\rangle \langle -1| +  C^\dagger_\uparrow \sigma^\mu C_\uparrow  \otimes |1\rangle \langle 1|  
\end{align*} 
where $C_\uparrow = \lvert \uparrow \rangle \langle \uparrow \rvert \cdot C$ and  $C_\downarrow = \lvert \downarrow \rangle \langle \downarrow \rvert \cdot C$ to unclutter notation and we used $T_+^\dagger = T_-$. More generally, the evolution of operator $|l\rangle \langle l |$ contains the linear combination of terms $|l\pm 1 \rangle \langle l\pm 1|$ that are orthonormal to each other. In what follow, it will prove important to represent the operators as 
\begin{align}\label{eq:op1op2}
	\mathcal{O}_0 &=
	\begin{gathered}
		\includegraphics[height=2cm]{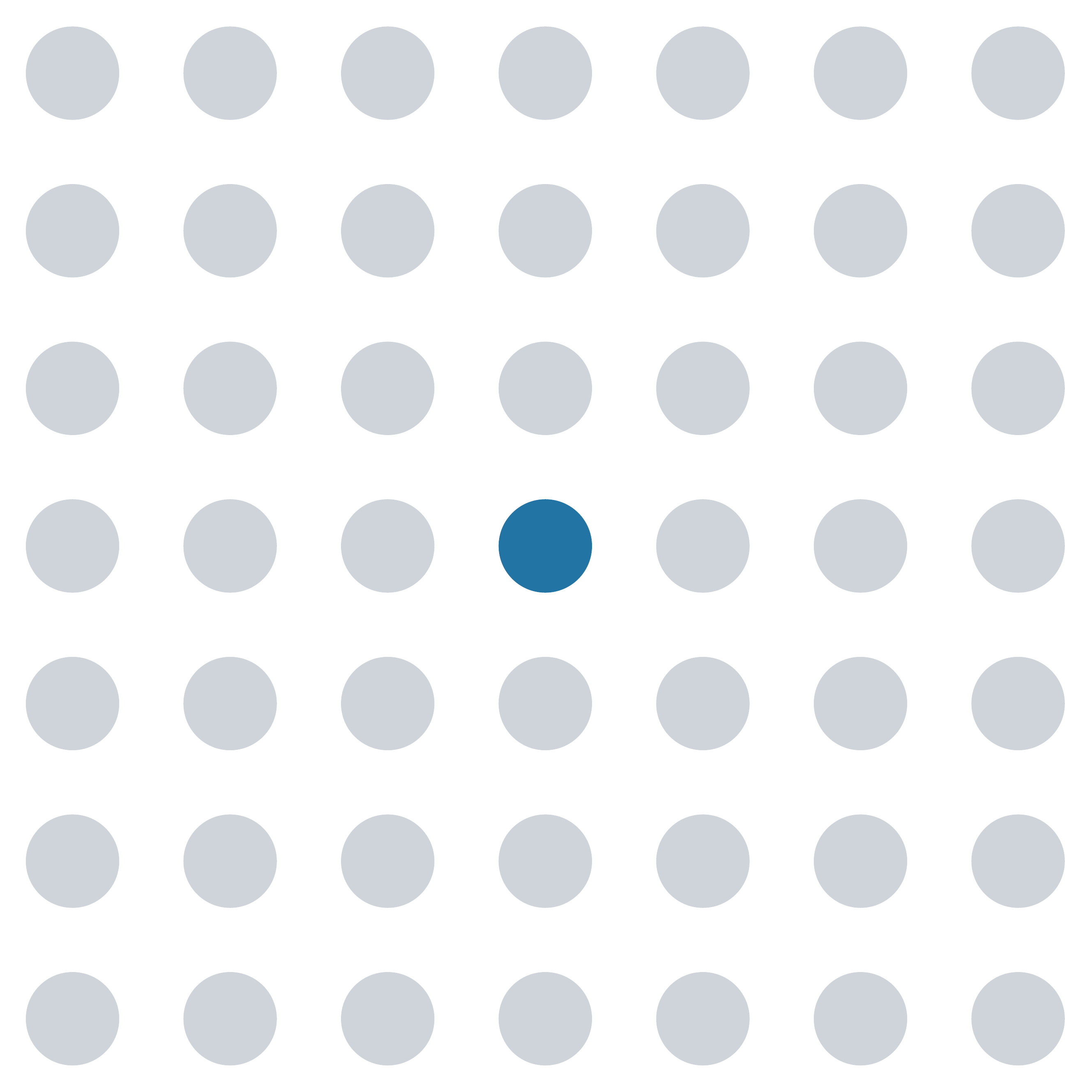}
	\end{gathered}  \qquad 
	\mathcal{O}_1 =
	\begin{gathered}
		\includegraphics[height=2cm]{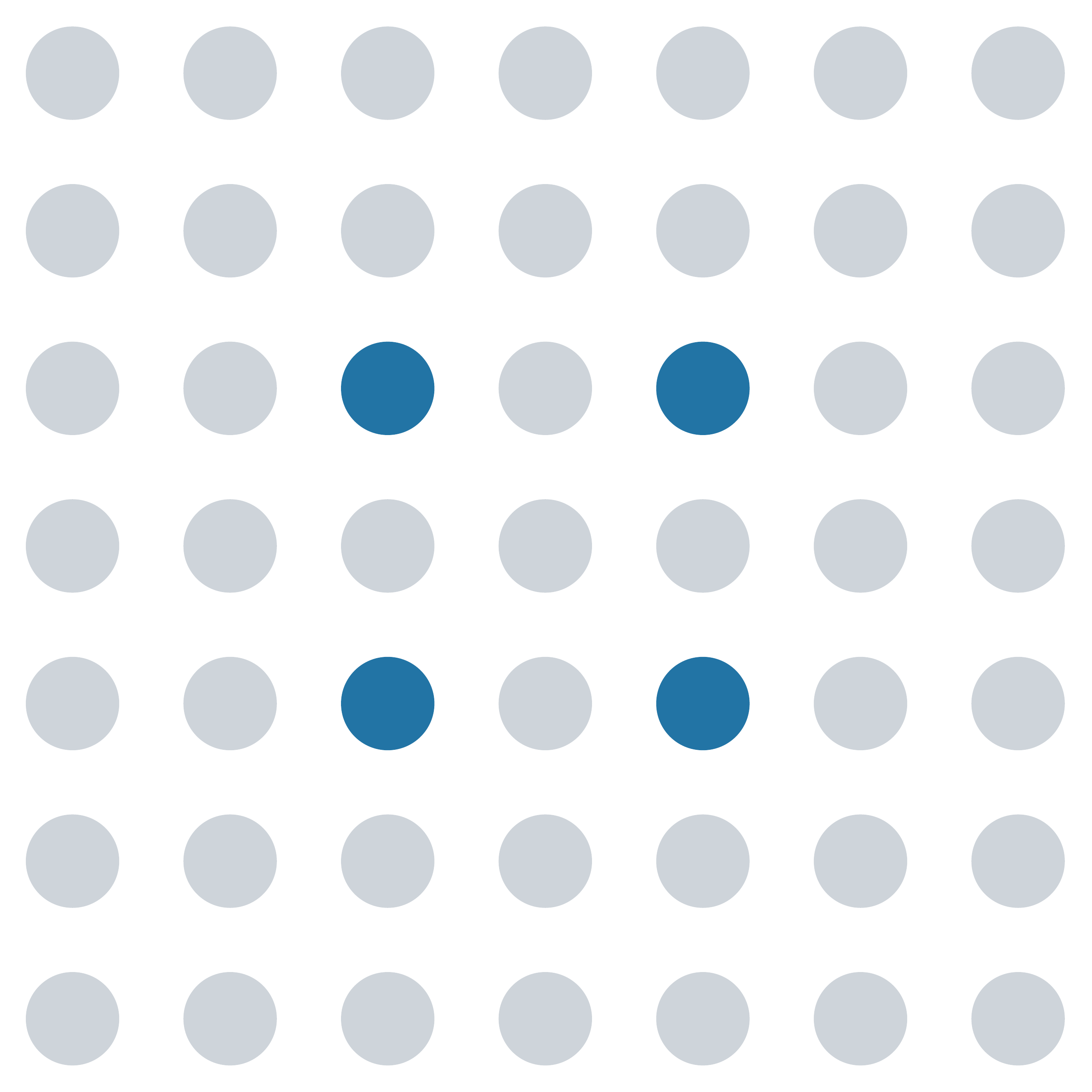}
	\end{gathered} 
\end{align}
where the blue dot represent the presence of term $P_{ij} = |i\rangle\langle  j|$ and multiple presence of blue dot represents the linear combination of these terms in operator expansion. The inner product between operators $\mathcal{O}_i$ and $\mathcal{O}_j$ depends on the expansion term corresponding to where dot color matches. The inner product between the operators in Eq.~\eqref{eq:op1op2} is zero i.e. the operators are orthonormal to each other since there are no common blue dots. At $t=2,3$, the operators are given by 
\begin{equation}\label{eq:O2}
	\mathcal{O}_2=
	\begin{gathered}
		\includegraphics[height=2cm]{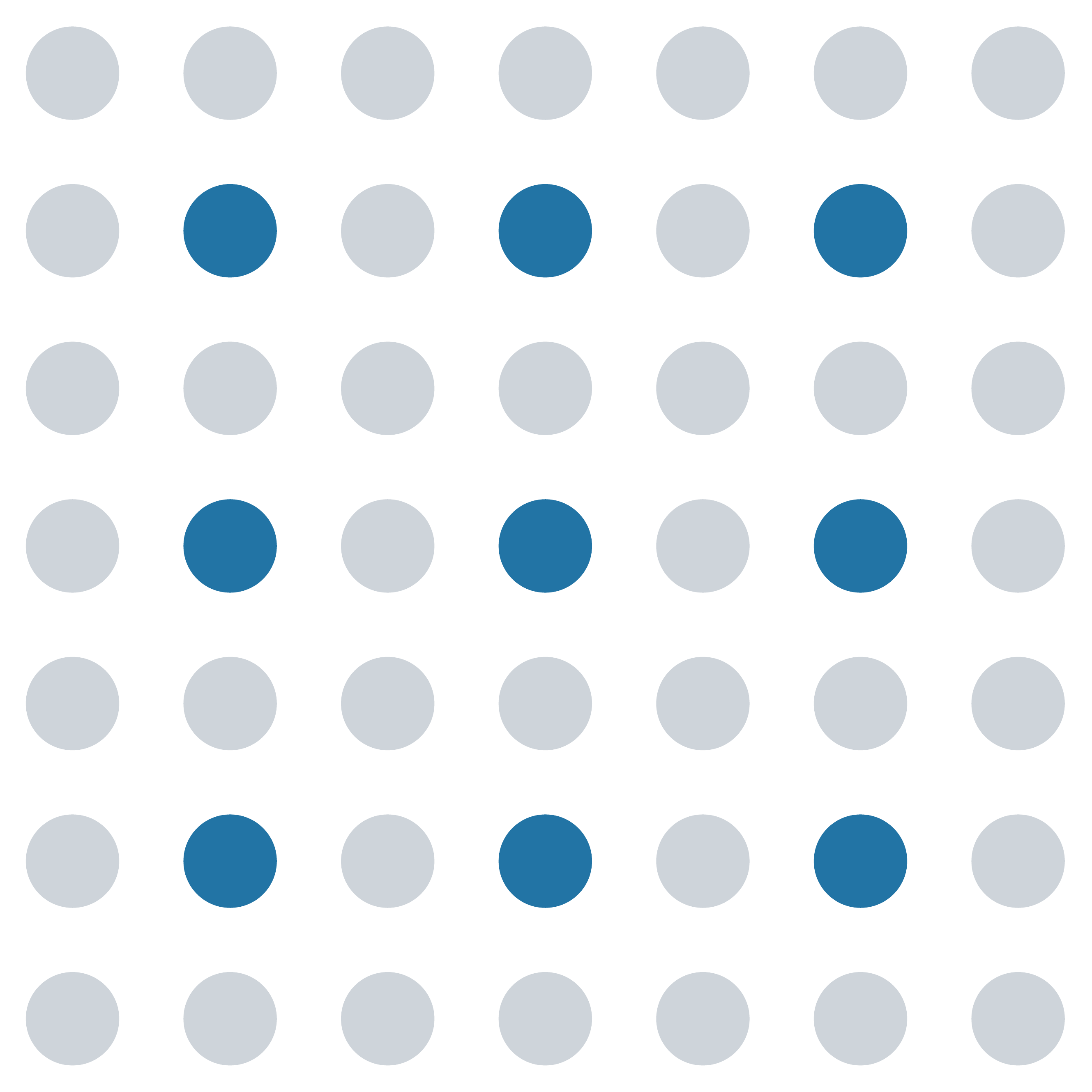}
	\end{gathered} \qquad  
	\mathcal{O}_3=
	\begin{gathered}
		\includegraphics[height=2cm]{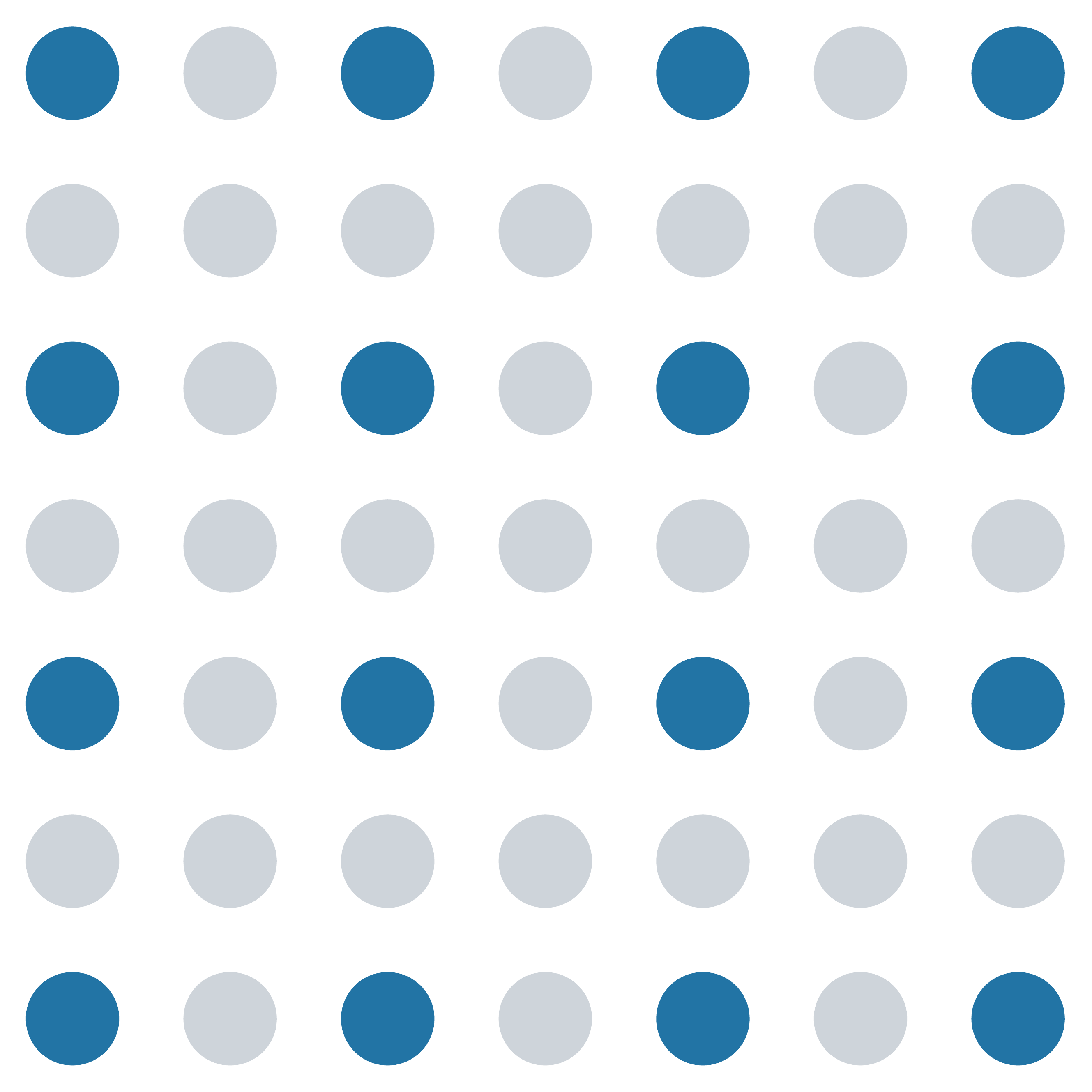}
	\end{gathered} 
\end{equation}
which shows $(\mathcal{O}_2|\mathcal{O}_1) = 0$ and $(\mathcal{O}_3| \mathcal{O}_2)  = (\mathcal{O}_3|\mathcal{O}_0) = 0$. In general, it follows that $(\mathcal{O}_t |\mathcal{O}_{t+1})  = 0$. If we define a matrix $\mathcal{A}$ such that $\mathcal{A}_{nm}= (\mathcal{O}_n|\mathcal{O}_m)$, it be such that it's odd off-diagonal terms are zero. We now prove that even off-diagonal terms are equal to each other. To show this consider,
\begin{align*}
	\mathcal{A}_{t,t-2}&\equiv (\mathcal{O}_t|\mathcal{O}_{t-2}) \propto \text{Tr}\left(\mathcal{O}^\dagger_t \mathcal{O}_{t-2}\right) \\
	&= \text{Tr} \left(U^\dagger \mathcal{O}^\dagger_{t-1} U \mathcal{O}_{t-2} \right) \\
	& = \text{Tr}\left(\mathcal{O}^\dagger_{t-1}\mathcal{O}_{t-3}\right) \equiv \mathcal{A}_{t-1,t-3}
\end{align*}
as required. In summary, we can write 
\begin{equation}
	\mathcal{A} = \begin{bmatrix}
		1& 0 & \mathcal{A}_{0,2} & 0 & \mathcal{A}_{0,4} & \cdots \\
		0 & 1 & 0 & \mathcal{A}_{0,2} & 0 & \ddots &  \\ 
		\mathcal{A}_{0,2} & 0& 1 & 0 & \mathcal{A}_{0,2} &\ddots \\
		0 & \mathcal{A}_{0,2} & 0 & 1 & 0 &  \ddots \\ 
		\vdots & \ddots & \ddots & \ddots & \ddots & \vdots  
	\end{bmatrix}
\end{equation}
Therefore, the krylov basis vector $|O_n)$ given by 
\begin{align*}
	|A_n) & = |\mathcal{O}_n) - \sum_{i<n} (O_i|\mathcal{O}_n) |O_i) \rightarrow |O_n) = \frac{|A_n)}{\lVert A_n\rVert} \\
	|A_0) &= |\mathcal{O}_0) \\
	|A_1) & = |\mathcal{O}_1) \\
	|A_2) &=  |\mathcal{O}_2) - \mathcal{A}_{02}   |\mathcal{O}_0) \rightarrow \lVert A_2\rVert = 1 - |\mathcal{A}_{02}|^2 \\
	|A_3 ) & =  |\mathcal{O}_3) - \mathcal{A}_{02}  |\mathcal{O}_1)     \rightarrow \lVert A_3\rVert = 1 - |\mathcal{A}_{02}|^2 \\
\end{align*}
	\begin{figure*}
	\centering
	\includegraphics[width=0.29\linewidth]{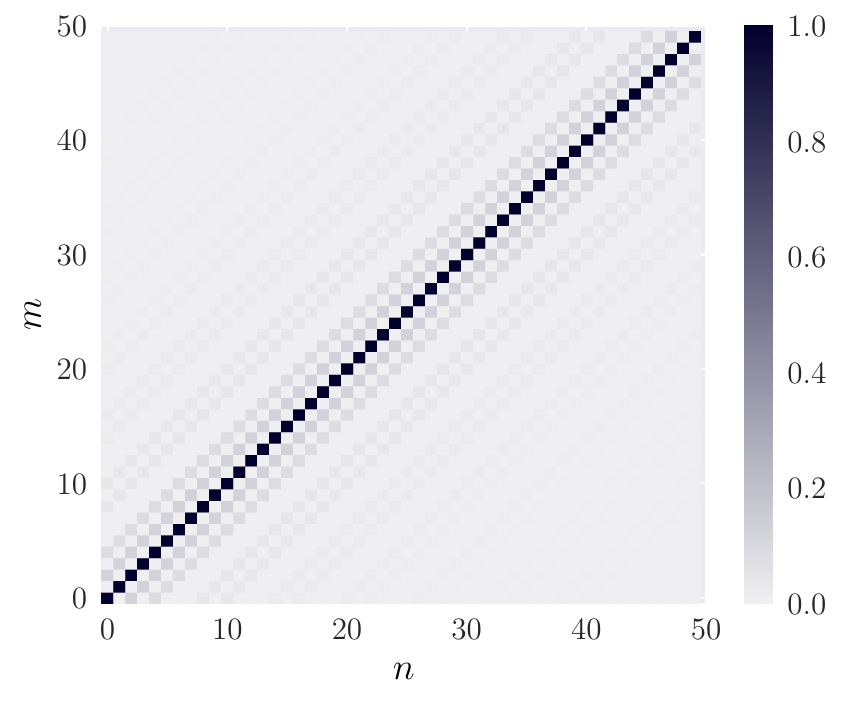}
	\includegraphics[width=0.35\linewidth]{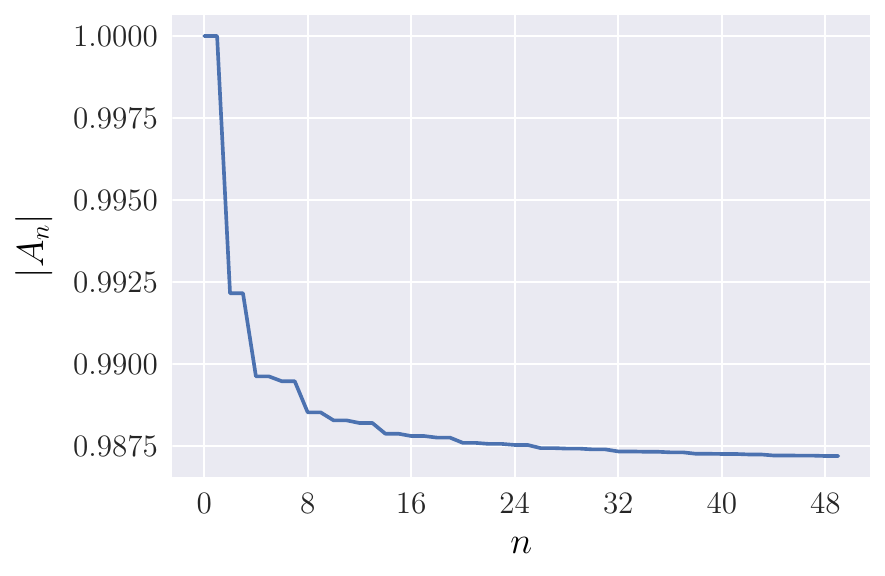}
	\includegraphics[width=0.33\linewidth]{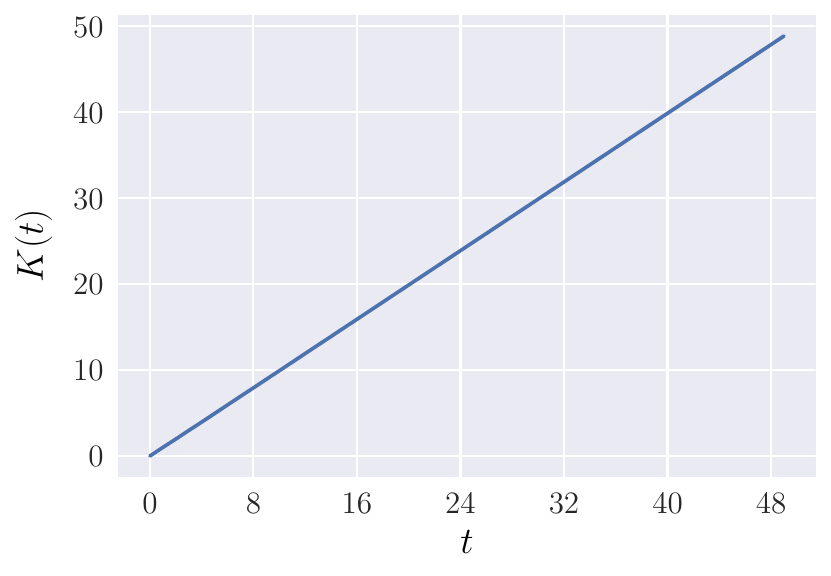}
	\caption{The numerical analysis for the Krylov complexity of discrete-time quantum walk with coin-angle $\theta = \pi/6$ and initial operator $\sigma^x\otimes |0\rangle \langle 0|$. \textbf{Left} The inner product matrix $\mathcal{A}_{nm} = (\mathcal{O}_n|\mathcal{O}_m)$ whose elements odd off-diagonal elements are $0$ while the even off-diagonal elements are equal. \textbf{Center} The norm of operator $\{A_n\}$ decays and then saturates to a value $\mathcal{O}(1)$. \textbf{Right} The K-complexity shows a linear growth. }
	\label{fig:anm}
\end{figure*}
As the dispersion in discrete-time quantum walk grows linearly\,\cite{chandrashekar2013disorder}, the operator support on a particular site decays over time. Hence, the overlap between the initial operator $\mathcal{O}_0$ and operator at any subsequent time $\mathcal{O}_{2t}$ also decays over time. In other words, $\mathcal{A}_{0,2t}\approx  0$, and therefore, $(\mathcal{O}_n|\mathcal{O}_m) \approx \delta_{nm}$. It follows that $\phi_{n,t} = (O_n|\mathcal{O}_t) \approx \delta_{n,t}$, hence the K-complexity given by
\begin{equation}
	K(t) = \sum_n n |\phi_n(t)|^2 \sim t
\end{equation} 
therefore, grows linearly. 
%In discrete-time evolving system, it is interesting to ask if the K-complexity cannot be superlinear? Since the evolution operator is used to generate the Krylov basis, each step results in at most a single Krylov basis vector --- Let's call is $|\mathcal{O}_n)$. The growth of the K-complexity would be maximum if this vector is orthonormal to $|O_i)$ for $i <n$ or $|O_n) = |\mathcal{O}_n)$. In this case, the wave function coefficients $\phi_{n,t} = \delta_{n,t}$ resulting in linear growth of K-complexity. 
In Fig.~\ref{fig:anm}, the numerical analysis is presented for K-complexity of the discrete-time quantum walk. The norm of operator $|A_n)$ saturates at a constant value close to $1$ showing that at late-time, the operators $|\mathcal{O}_n)$ are approximately orthogonal to each-other.  Therefore, the K-complexity obeys a linear growth. 

\begin{figure}
	\centering
	\includegraphics[width=0.49\linewidth]{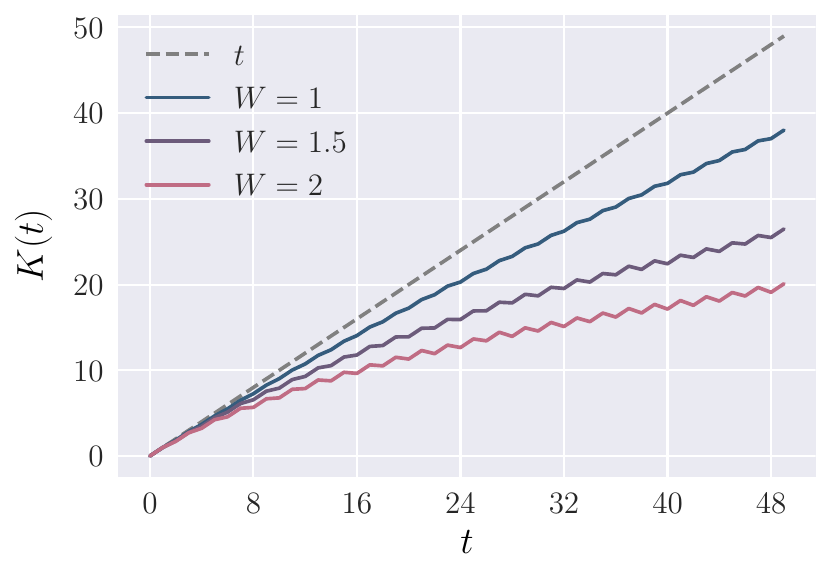}
	\includegraphics[width=0.49\linewidth]{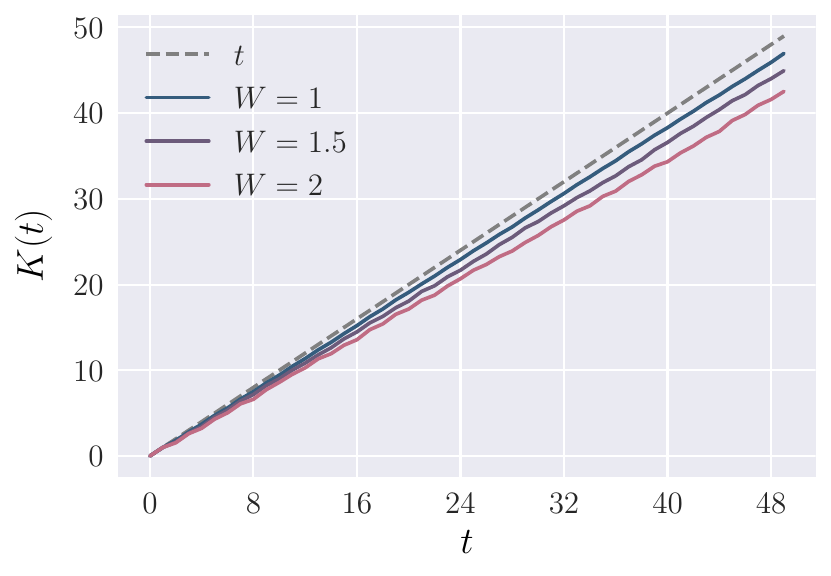}
	\caption{The K-complexity calculated for discrete-time quantum walk in the presence of \textbf{Left}: spatial \textbf{Right}: temporal disorder with varying disorder strength $W$. The disorder is taken over 500 copies. The dotted lines show the curve corresponding to $y(t) = t$.}
	\label{fig:Kdisorder}
\end{figure}

As we seen in Sec.~\ref{subsec:OTOC}, the introduction of disorder leads to slowdown in information propagation. The K-complexity shows the similar localized behavior as OTOC in which it deviates from linear growth to suppress power law $\sim t^{1/\delta}$ with $\delta >1$. In this case, the amplitude $\mathcal{A}_{0,2n}$ goes to zero for large $n$ due to localization at $n = 0$. Therefore, the expansion coefficient $\phi_{n,t}  = (O_n|\mathcal{O}_t)$ is non-zero for small $n$ which results in suppressed growth in complexity. In Fig.~\ref{fig:Kdisorder}, the K-complexity is calculated under temporal and spatial disorder in discrete-time quantum walk for increasing disorder strength $W$. As with the OTOC, the suppression in K-complexity growth is smaller in temporal case compared to spatial as result of strong localization (or AL) in the later case. 

\section{Conclusion}

In summary, the discrete-time quantum walks are quantum model which can be used to simulate a large number of phenomenon from quantum many-body physics as well as for construction of quantum algorithms. There experimental implementation on wide number of platform makes them suitable to study theoretical ideas. In this article, we study information scrambling in discrete-time quantum walk using the out-of-time correlators and K-complexity as probe. The OTOC features the ``shell-like'' behavior, in which, at long-time limit, it goes to zero indicating no scrambling of the operator. The introduction of disorder (spatial or temporal) results in slowdown of information scrambling where the shape of light-cone confines up to localization length.  The K-complexity shows a linear growth which ties to the fact that the operator at any time is approximately orthonormal to all operators at previous times. The disorder suppresses the K-complexity growth resulting in its sub-linear behavior. While the spatial disorder results in strong localization, the temporal disorder results in weak localization which is apparent from the behavior of both OTOC as well as K-complexity.

We would like to understand the effect of boundary conditions. In this work, we focused discrete-time quantum walk in infinite-dimensional 1D lattice, but from experimental point of view, it may be interesting to see the late-time behavior of these quantities $t>L$. In Ref.\,\cite{PhysRevB.107.144306}, the information scrambling was studied on Clifford quantum cellular automata (QCA). These systems shown to break ergodicity, i.e., they exhibit quantum scarring. It was further shown that such a system could exhibit classical dynamics in some semi-classical limit. While the discrete-time quantum walks are not the same as QCA, but they can be regarded as the dynamics of the one-particle sector of a QCA\,\cite{Farrelly2020reviewofquantum}. Therefore, it will be interesting to see if the connection between the two results can be made more direct. In this direction, one should note that Clifford QCAs model studied in ref.\,\cite{PhysRevB.107.144306} also exhibits linear growth in K-complexity. It follows from the evolution of operator under Clifford QCAs in which a string of pauli operators maps to another. Since the pauli operator form a orthonormal basis, each evolution step generate a new basis element resulting in linear growth. While for infinite-dimensional case, this growth will persist forever due to the infinite dimensional Hilbert space, in case of finite-dimensional lattice, the growth will stop as the operator get backs to its initial state. Therefore, the linear growth in K-complexity follows by a sudden decay to zero followed by repetitive behavior which is the refinancing of recurrence dynamics. 

In this work, we formulated the K-complexity for the system with discrete-time evolution. Therefore, it opens up a way to consider more interesting many-body systems with discrete-time evolution. One such example could be random unitary circuits (RUCs)\,\cite{annurev-conmatphys-031720-030658}, which has been an active area of research for the past several years. RUCs have shed light on longstanding questions about thermalization and chaos and on the underlying universal dynamics of quantum information and entanglement\,\cite{PhysRevX.7.031016}. While several works have explored the dynamics of OTOC in a number of variants of RUCs\,\cite{PhysRevX.8.021014,PhysRevLett.131.220404}, the study of K-complexity can further help understand post-scrambling-time behavior.  

From the point of view of discrete-time quantum walks, which are the basis of many quantum algorithms, the study of information scrambling could provide the basis for designing algorithms that can efficiently explore the Hilbert spaces. In this context, the K-complexity which describe the delocalization of the operator in Hilbert space may be useful. Overall, the study of scrambling in both discrete-time quantum walks as well as in more generic systems with discrete-time evolution paves the way toward new physics in quantum many-body physics.

\smallskip
\textit{Acknowledgement:} We wish to thank Aranya Bhattacharya for various useful discussions and comments about this and related works. We thank the anonymous referees for their thoughtful comments and constructive criticism.

\vfill

\smallskip
\section{References}

\bibliographystyle{apsrev4-2}
\bibliography{biblio}

\end{document}